\documentclass{ws-procs9x6}
\usepackage{amsmath}
\usepackage{amssymb}
\usepackage{amsfonts}
\usepackage{latexsym}
\def\one{1\hskip -.37em 1}     

\begin{document}

\title{On the Road Towards\\
the Quantum Geometer's Universe:\\
an Introduction to Four-Dimensional\\
Supersymmetric Quantum Field Theory}

\author{Jan GOVAERTS}

\address{Institute of Nuclear Physics, Catholic University of Louvain\\
2, Chemin du Cyclotron, B-1348 Louvain-la-Neuve, Belgium\\
E-mail: govaerts@fynu.ucl.ac.be}


\maketitle

\abstracts{
This brief set of notes presents a modest introduction
to the basic features entering the construction of
supersymmetric quantum field theories in four-dimensional Minkowski 
spacetime, building a bridge from similar lectures presented at 
a previous Workshop of this series, and reaching only at the doorstep 
of the full edifice of such theories.
}

\section{Introduction}
\label{Sect1}

The organisers of the third edition of the COPROMAPH Workshops
had thought it worthwhile to have the second series of lectures
during the week-long meeting dedicated to an introduction to
supersymmetric quantum field theories. An internationally renowned
expert in the field had been invited, and was to deliver
the course. Unfortunately, at the last minute
fate had it decided otherwise, depriving the participants of what 
would have been an introduction to the subject of outstanding quality.
The present author was finally found to be on hand, without being able 
to do full justice to the wide relevance of the topic, ranging from 
pure mathematics and topology to particle physics phenomenology at its 
utmost best in anticipation of the running of the LHC at CERN by 2007.

Fate had it also that the same author had already delivered a similar series
of lectures at the previous edition of the COPROMAPH Workshops,\cite{GovCOPRO2}
which in broad brush strokes attempted to paint with vivid colours
the fundamental principles of XX$^{\rm th}$ century physics,
underlying all the basic conceptual progresses having led to 
the relativistic quantum gauge field theory and classical general 
relativity frameworks for the present description of all known forms 
of elementary matter constituents and their fundamental interactions,
as inscribed in the Standard Model of particle physics and Einstein's
classical theory of general relativity. At the same time, a few of
the doors onto the roads winding deep into the unchartered territories 
of the physics that must lie well beyond were also opened. It is thus
all too fitting that we get the opportunity to trace together 
a few steps onto one of these roads, in the embodiement of a Minkowski 
spacetime structure extended into a superspace including now 
also anticommuting coordinates in addition to the usual commuting 
spacetime ones. We are truly embarking on a journey onto the roads
leading towards the quantum geometer's universe! Even if only by marking the
path by a few white and precious pebbles to guide us into the unknown
territory when the time will have come for more solitary explorations 
of one's own in the composition, with a definite African beat,
of the music scores of the unfinished symphony of XXI$^{\rm st}$ century 
physics.\cite{GovCOPRO2}

Even though none are based on actual experimental facts, there exist
a series of theoretical and conceptual
motivations for considering supersymmetric extensions of
ordinary Yang--Mills theories in the quest for a fundamental
unification. Spacetime supersymmetry is a symmetry that exchanges
particles of integer --- bosons --- and 
half-integer --- fermions --- spin,\footnote{The lectures delivered at
COPROMAPH2 did not deal with field theories associated
to fermionic degrees of freedom described using Grassmann odd
variables, and considered only bosonic theories.\cite{GovCOPRO2} 
Quantised fermionic field theories are briefly dealt with
in Sec.~\ref{Sec3}.} enforcing specific relations between the properties 
and couplings of each class of particles, when supersymmetry remains
manifest in the spectrum of the system. In particular, since for what
concerns ultra-violet (UV) short-distance divergences of quantum field theories
in four-dimensional Minkowski spacetime fermionic fields are less ill-behaved
than bosonic fields (namely, in terms of a cut-off in energy, divergences
in fermionic loop amplitudes are usually only logarithmically divergent
whereas those of bosonic loops are quadratically divergent), one should
expect that in the presence of manifest supersymmetry, UV divergences
should be better tamed for bosonic fields, being reduced to a logarithmic
behaviour only as in the fermionic sector (this has important consequences
which we shall not delve into here). Another aspect is that
within the context of superstring and M-theory\cite{Strings} with bosonic 
and fermionic states, quantum consistency is ensured provided 
supersymmetries are restricting the dynamics. In this sense, the existence 
of supersymmetry at some stage of unification beyond the Standard Model 
is often considered to be a natural prediction of M-theory.

Besides such physics motivations just hinted at, supersymmetry has 
also proved to be of great value in mathematical physics, in the 
understanding of nonperturbative phenomena in quantum field theories 
and M-theory,\cite{Witten1,MATH} and for uncovering deep connections 
between different fields of pure ma\-the\-ma\-tics. The algebraic structures 
associated to Grassmann graded algebras are powerful tools with which 
to explore new limits in the concepts of geometry, topology and 
algebra.\cite{MATH} One cannot help but 
feel that a great opportunity would be missed if tomorrow's quantum geometry
would not make any use of supersymmetric algebraic structures.

Since its discovery in the early 1970's,\cite{SUSY1970,WZ} applications 
of supersymmetry have been developed in such a diversity of directions 
and in so large a variety of fields of physics and mathematics, that 
it is impossible to do any justice to all that work in the span of any 
set of lectures, let alone only a few. Our aim here will thus be very modest. 
Namely, starting from the contents of the previous lecture 
notes,\cite{GovCOPRO2} build a bridge reaching the entry roads and 
the shores towards supersymmetric 
field theories and the fundamental concepts entering their construction.
Not that the lectures delivered at the Workshop did not discuss
the general superfield approach over superspace as the most
efficient and transparent techniques for such constructions in the case 
of $\mathcal N=1$ supersymmetry, but the latter material being so widely 
and in such detailed form available from the literature, it is felt that rather
a detailed introduction to the topics missing from Ref.~\refcite{GovCOPRO2} 
but necessary to understand supersymmetric field theories is of greater 
use and interest to most readers of this Proceedings volume. With 
these notes, our aim is thus to equip any interested reader with 
a few handy concepts and tools to be added to the backpack to be carried 
on his/her explorer's journey towards the quantum geometer's universe 
of XXI$^{\rm st}$ century physics, in search of the new principle beyond
the symmetry principle of XX$^{\rm th}$ century physics.\cite{GovCOPRO2}

Also by lack of space and time, even of the anticommuting type
if the world happens to be supersymmetric indeed, we shall thus stop
short of discussing explicitly any supersymmetric field theory
in 4-dimensional Minkowski spacetime, even the simplest example
of the $\mathcal N=1$ Wess-Zumino model\cite{WZ} that may be constructed
using the hand-made tools of an amateur artist-composer in the art of
supersymmetries. From where we shall leave the subject in these notes,
further study could branch off into a variety of
directions of wide ranging applications, beginning with general
supersymmetric quantum mechanics and the general
superspace and superfield techniques for $\mathcal N=1$ and $\mathcal N=2$
supersymmetric field theories with Yang--Mills internal gauge symmetries 
and the associated Higgs mechanism of gauge symmetry breaking, to further 
encompass the search for new physics at the LHC through the construction of 
supersymmetric extensions of the Standard Model, or also reaching towards 
the duality properties of supersymmetric Yang--Mills and M-theory, mirror 
geometry, topological string and quantum field theories,\cite{GovTQFT} 
etc., to name just a few examples.\cite{Strings,Witten1,MATH}

Let us thus point out a few standard textbooks and lectures
for large and diversified accounts of these classes of
theories and more complete references to the original literature. 
Some important such material is listed in Refs.~\refcite{WB}
and \refcite{SUSYQM}.
In particular, the lectures delivered at the Workshop were
to a si\-gni\-fi\-cant degree inspired by the contents of Ref.~\refcite{Deren}. 
Any further search through the SPIRES databasis 
({\tt http://www.slac.stanford.edu/spires/hep/}; 
UK mirror: {\tt http://www-spires.dur.ac.uk/spires/hep/}) will quickly 
uncover many more useful reviews.

In Sec.~\ref{Sec2}, we briefly recall the basic facts of relativistic
quantum field theory for bosonic degrees of freedom, discussed at greater
length in Ref.~\refcite{GovCOPRO2}, in order to explain why such systems
are the natural framework for describing relativistic quantum 
point-particles. The same considerations are then developed 
in Sec.~\ref{Sec3} in the case of fermionic degrees of freedom associated
to particles of half-integer spin, based on a discussion of the 
theory of finite dimensional representations
of the Lorentz group, leading in particular to
the free Dirac equation for the description of spin 1/2 particles
without interactions. Section~\ref{Sec4} then considers, as a simple
introductory illustration of some facts essential and generic to
supersymmetric field theories, and much in the same spirit as
that of the discussion in Sec.~\ref{Sec2}, the $\mathcal N=1$ 
supersymmetric harmonic oscillator which already displays quite 
a number of interesting properties.
Section~\ref{Sec5} then concludes with a series of final remarks
related to the actual construction of supersymmetric field theories
based on the general concepts of the Lie symmetry algebraic structures 
inherent to such relativistic invariant quantum field theories and their
manifest realisations through specific choices of field content, indeed the
underlying theme to both these lectures and the previous ones.\cite{GovCOPRO2}

\section{Basics of Quantum Field Theory:
A Compendium for Scalar Fields}
\label{Sec2}

Within a relativistic classical framework,\footnote{Throughout most
of these notes, units such that $c=1=\hbar$ are used.} material reality 
consists, on the one hand, of dynamical fields, and on the other hand, 
of point-particles. Fields act on particles through forces that they 
develop, such as the Lorentz force of the electromagnetic field 
for charged particles, while particles react back onto the fields 
being sources for the latter, for instance through the charge and 
current density sources of the electromagnetic field in Maxwell's equations
(the same characterisation applies to the gravitational field equations of
general relativity). This dichotomic distinction between matter and 
radiation is unified in a dual form when considering a relativistic 
quantum framework. Indeed, it then turns out that particles are nothing else 
than the quanta, {\it i.e.\/}, the quantum states of definite energy,
momentum and spin, of quantum fields. Particles and fields are just 
the two complementary aspects of the quantum relativistic world of 
point-particles. All electrons, for example, are identical, being 
quanta of a single electron field filling all of spacetime. To each 
distinct species of particle corresponds a field, and vice-versa. 
This, in a word, is the essence of quantum field theory: the natural 
framework for a description of relativistic quantum point-particles, 
explaining their corpuscular properties when detected in energy-momentum 
eigenstates and their wave behaviour when considering their spacetime 
propagation. Let us briefly express these points in a somewhat 
more mathematical setting.

\subsection{Particles and Fields}
\label{Sec2.1}

A free relativistic field may be seen to correspond to an infinite collection
of harmonic oscillators sitting at each point in space, and coupled to
one another through a nearest-neighbour term in the action of the field's
dynamics. Let us first recall a few basic facts about the one-dimensional 
harmonic oscillator. Its dynamics derives through the variational principle
from the action
\begin{equation}
S[q]=\int dt\,m\left[\frac{1}{2}\left(\frac{dq}{dt}\right)^2-
\frac{1}{2}\omega^2q^2\right]\ ,
\label{eq:SHO}
\end{equation}
where $q(t)\in\mathbb R$ is the configuration space of the system.
How to perform the standard canonical operator quantisation of this
system is well known,\cite{GovCOPRO2} leading, in the Heisenberg picture,
to the following quantum operator re\-pre\-sen\-ta\-tion,
\begin{equation}
\hat{q}(t)=\sqrt{\frac{\hbar}{2m\omega}}\,
\left[a\,e^{-i\omega t}\,+\,a^\dagger\,e^{i\omega t}\right]\ ,
\label{eq:solHO}
\end{equation}
obeying the operator equation of motion
\begin{equation}
\left[\,\frac{d^2}{dt^2}\ +\ \omega^2\,\right]\,\hat{q}(t)=0\ .
\label{eq:ELHO}
\end{equation}
Here, $a$ and $a^\dagger$ are the annihilation and creation operators
for the quanta of the system (they are complex conjugate integration
constants at the classical level), obeying the Fock space algebra
\begin{equation}
[a,a^\dagger]=\one\ .
\end{equation}
The quantum Hamiltonian $\hat{H}=\hbar\omega(a^\dagger a+1/2)$ is
diagonal in the Fock state basis, constructed as follows for all natural
numbers $n=0,1,2,\cdots$,
\begin{equation}
|n\rangle=\frac{1}{\sqrt{n!}}(a^\dagger)^n|0\rangle\ \ ,\ \ 
\langle n|m\rangle=\delta_{nm}\ \ ,\ \ 
a|0\rangle=0\ \ ,\ \ \hat{H}|n\rangle=\hbar\omega(n+\frac{1}{2})|n\rangle\ .
\end{equation}
The physical interpretation is that the state $|0\rangle$ defines
the ground state or vacuum of the quantum oscillator, with the
discrete set of states $|n\rangle$ ($n=1,2,\cdots$) corresponding
to excitations of the oscillator with $n$ quanta each contributing
an energy $\hbar\omega$ on top of the vacuum quantum energy $\hbar\omega/2$
due to so-called vacuum quantum fluctuations. In particular, the operators
$a$ and $a^\dagger$ are ladder operators between the Fock states,
\begin{equation}
a|n\rangle=\sqrt{n}|n\rangle\ \ ,\ \ 
a^\dagger|n\rangle=\sqrt{n+1}|n+1\rangle\ \ ,\ \ 
a^\dagger a|n\rangle=n|n\rangle\ .
\end{equation}

Thus, here we have a mathematical framework in which the quantisation
of a configuration space $q(t)\in\mathbb R$ leads to an algebra of
quantum operators representing the creation and annihilation of
energy eigenstates. In order to describe the dynamics of relativistic
quantum point-particles which likewise, as observed in experiments,
may be created and annihilated, we shall borrow a similar
mathematical framework. Since the harmonic oscillator is a system
invariant under translations in time, according to Noether's 
theorem\cite{GovBook} there must exist a conserved quantity associated 
to this continuous symmetry whose value coincides with the energy 
of the system, namely its Hamiltonian. In the case of relativistic 
particles defined over
Minkowski spacetime,\footnote{Our choice of Minkowski spacetime
metric signature is such that $\eta_{\mu\nu}={\rm diag}\,(+---)$ 
with $\mu,\nu=0,1,2,D-1$ with $D=4$.} invariance under spacetime 
translations implies the existence
of conserved quantities associated to these symmetries, namely the particle's
total energy and momentum, which we shall denote $k^\mu=(k^0,\vec{k}\,)$
with $k^0=\omega(\vec{k}\,)=\sqrt{\vec{k}\,^2+m^2}$, $m$ being the particle's
mass. Consequently, let us introduce the annihilation and creation
operators, $a(\vec{k}\,)$ and $a^\dagger(\vec{k}\,)$, respectively,
for particles of given momentum $\vec{k}$ and energy $\omega(\vec{k}\,)$, 
and obeying the commutation relations\footnote{Only the nonvanishing 
commutators are given. The choice of normalisation is made such that 
the momentum integration measure in the mode decomposition of the fields 
later on is Lorentz invariant.\cite{GovCOPRO2} This choice also implies that
particle states are normalised in a Lorentz covariant manner.}
\begin{equation}
\left[a(\vec{k}\,),a^\dagger(\vec{\ell}\,)\right]=
(2\pi)^3\,2\omega(\vec{k}\,)\,\delta^{(3)}(\vec{k}-\vec{\ell}\,)\ .
\end{equation}
For instance, 1-particle states are thus constructed as
\begin{equation}
|\vec{k}\rangle=a^\dagger(\vec{k}\,)|0\rangle\ \ ,\ \ 
\langle\vec{k}|\vec{\ell}\,\rangle=(2\pi)^3 2\omega(\vec{k}\,)\,
\delta^{(3)}(\vec{k}-\vec{\ell}\,)\ ,
\end{equation}
$|0\rangle$ being the Fock vacuum of the system.

In order to identify the actual configuration space of the system
that is being considered, by analogy with (\ref{eq:solHO}), let us
construct the following quantum operator in the Heisenberg picture,
\begin{equation}
\hat{\phi}(x^\mu)=\int\frac{d\vec{k}}{(2\pi)^32\omega(\vec{k}\,)}
\left[a(\vec{k}\,)\,e^{-ik\cdot x}\,+\,
a^\dagger(\vec{k}\,)\,e^{ik\cdot x}\,\right]\ ,
\end{equation}
where the inner product in the plane wave factors is defined to be given
by $k\cdot x=\omega(\vec{k}\,)x^0-\vec{k}\cdot\vec{x}$, thus with the
on-shell energy value $k^0=\omega(\vec{k}\,)$. Note that in
comparison to (\ref{eq:solHO}), the plane wave factors, corresponding
to positive- and negative-frequency components of the wave equation and
involving only a time dependence in the case of the harmonic oscillator, have
now been extended to the spacetime dependent
Lorentz invariant quantity $k\cdot x$.
The requirement of a relativistic covariant description of quantum
point-particles being created and annihilated requires such an
extension of the plane wave contributions. Furthermore, the
operator $\hat{\phi}(x^\mu)$ obeys the quantum equation of motion
\begin{equation}
\left[\,\frac{\partial^2}{\partial t^2}\,-\,
\vec{\nabla}^2\,+\,m^2\,\right]\,\hat{\phi}(x^\mu)=0\ ,
\end{equation}
in which one recognises of course the Klein--Gordon equation,
deriving also from the action
\begin{equation}
S[\phi]=\int dt\int_{(\infty)} d^3\vec{x}\,
\left[\frac{1}{2}\left(\frac{\partial\phi}{\partial t}\right)^2-
\frac{1}{2}\left(\vec{\nabla}\phi\right)^2-\frac{1}{2}m^2\phi^2\right]\ .
\label{eq:KGaction}
\end{equation}

In other words, such a framework for the description of relativistic
quantum point-particles and their creation and annihilation naturally
leads to a relativistic quantum field theory. The configuration space
of such a system is that of the relativistic real scalar field $\phi(x^\mu)$.
In the quantum world, configurations of this field 
are observed through its energy and
momentum eigenstates --- thanks to the invariance under spacetime translations
of the Klein--Gordon action --- which are nothing else than the particle
quanta of the field. To any relativistic quantum field one associates
relativistic quantum point-particles, and to any ensemble of indistinguishable
relativistic quantum point-particles one associates a relativistic quantum 
field. Fields and particles are only two dual aspects in a relativistic 
quantum universe whose basic ``constituents'' are dynamical fields.  
Note that relativistic covariance, which forced the extension to the
Lorentz invariant $k\cdot x$ in the plane wave contributions,
also explains the appearance of the gradient terms $\vec{\nabla}\phi$
in the Klein--Gordon wave equation and action. Had this term been
absent, indeed the field $\phi(x^\mu)$ would have described simply
an infinite collection of harmonic oscillators 
$q_{\vec{x}}(t)=\phi(t,\vec{x}\,)$ each fixed at each of the points
in space, and oscillating independently of one another. However,
the gradient term introduces a specific nearest-neighbour coupling
between these oscillators, such that any disturbance set-up in any one of
them will quickly spread throughout space in a wave-like manner because
of the gradient coupling between adjacent oscillators. These linear
waves are characterised by their wave-number vector $\vec{k}$ and frequency
$\omega(\vec{k}\,)$, which, at the quantum level, are identified
with the quanta's or particles' conserved momentum and energy. Indeed,
because of Noether's theorem associated to the invariance under spacetime
translations, the total energy and momentum of the field take these
values for the 1-particle quantum states $|\vec{k}\rangle$. The
system is also invariant under the full Lorentz group of spacetime
(pseudo)rotations, hence leading also to further conserved quantum
numbers of the field and its quanta associated to their spin. In the
present instance, since the field $\phi(x^\mu)$ transforms as a
scalar under the Lorentz group (namely, it is left invariant), the quanta
of such a real scalar field carry zero spin.

The above argument thus explains why relativistic quantum field
theory provides the natural framework for the description of
relativistic quantum point-particles that may be annihilated and
created. An explicit canonical quantisation starting from
the classical Klein--Gordon action (\ref{eq:KGaction}) of course recovers
all the above results, simply by applying the usual rules of
quantum mechanics to this system of degrees of freedom 
$q_{\vec{x}}(t)=\phi(t,\vec{x}\,)$.\cite{GovCOPRO2}

Furthermore, it is also possible to set up a perturbation
expansion for the introduction of spacetime local interactions
between such fields or with themselves (simply by adding to the
Klein--Gordon Lagrangian density higher order products of
the fields at each point in spacetime, thus preserving spacetime
locality and causality), and the systematic calculation, through 
Feynman diagrams and the corresponding Feynman rules,
of matrix elements of the scattering S-matrix.\cite{GovCOPRO2} Hence finally,
decay rates and cross-sections for processes occuring between the
quanta associated to such fields may be evaluated, at least through
perturbation theory, starting from any given quantum field theory 
extended to include also interactions.

As discussed in Ref.~\refcite{GovCOPRO2}, it is at this stage that
the short-distance UV divergences appear in loop amplitudes, for
which the renormalisation programme has been designed.\cite{QFT} Large classes
of renormalisable, {\it i.e.\/}, theories for which specific predictions
may be made, have thereby been identified, and they all fall within
the class of Yang--Mills theories of local gauge interactions extended
in different manners and involving particles of spin 0 and~1/2 for
the matter fields, and of spin 1 for the gauge fields associated to
the gauge interactions. These are the basic concepts going into the
construction of the Standard Model of the strong and electroweak interactions,
based on the gauge symmetry group 
SU(3)$_{\rm C}\times$SU(2)$_{\rm L}\times$U(1)$_{\rm Y}$.
A brief discussion of Yang--Mills theories,
the Higgs mechanism of spontaneous symmetry breaking and the
generation of mass is available in Ref.~\refcite{GovCOPRO2}.

All the above may readily be extended to collections of real
scalar fields. When further internal symmetries appear,\footnote{For
example, taking two real scalar fields of identical mass defines
a system with a global SO(2)=U(1) symmetry. The associated conserved quantum
number thus distinguishes quanta according to their U(1) quantum number,
taking a value either $(+1)$ for certain quanta --- particles --- or
$(-1)$ for other quanta --- antiparticles ---, all sharing otherwise
the same kinematical and spacetime properties such as mass and spin.
The existence of matter and antimatter is thus a natural outcome of
relativistic quantum field theory, extended to complex valued fields.
Indeed, the two mass-degenerate real scalar fields combine into a
single complex scalar field, invariant under any global, namely
spacetime independent transformation of its phase.} additional internal 
quantum numbers exist, by virtue of Noether's theorem, and particle 
quanta may then be classified according to specific linear representations 
of that internal symmetry group when realised in the Wigner 
mode.\footnote{This is no longer the case if the symmetry is realised 
in the Goldstone mode, namely when it is spontaneously broken by the 
vacuum which is then not invariant under the action of the 
symmetry.\cite{GovCOPRO2}} It thus proves very efficient to base 
the consideration of the construction of relativistic quantum field 
theories, of which the particle quanta carry collections of conserved
quantum numbers and specific interactions governed by the associated
symmetries, on a Lagrangian formulation, since symmetries
of the dynamics are then made manifest, readily leading to the
identification of the conserved quantities through the Noether
theorem. Thus when turning to the construction of field theories
possessing the invariance under supersymmetry transformations, the
analysis will be performed directly in terms of the Lagrangian density
once the field content is specified.

\subsection{Spacetime Symmetries}
\label{Sec2.2}

Since supersymmetry relates particles of integer and half-integer
spin, it is a symmetry that intertwines with the spacetime
symmetry of the full Poincar\'e group in Minkowski spacetime.
It is thus important that we first understand the basics of
the Poincar\'e group algebra, involved in the contruction of
any relativistic quantum field theory over Minkowski spacetime.

Acting on the spacetime coordinates, the ISO(1,3) Poincar\'e
group (in a 4-dimensional Minkowski spacetime)
is defined by the transformations
\begin{equation}
{x'}^\mu={\Lambda^\mu}_\nu\,x^\nu\ +\ a^\mu\ ,
\end{equation}
where the 4-vector $a^\mu$ represents a constant spacetime translation,
and ${\Lambda^\mu}_\nu$ a constant SO(1,3) Lorentz (pseudo)rotation
leaving invariant the Minkowski metric,
\begin{equation}
\eta_{\rho\sigma}{\Lambda^\rho}_\mu\,{\Lambda^\sigma}_\nu=
\eta_{\mu\nu}\ .
\end{equation}
These transformations also act on the field content of a given
theory. In the case of scalar fields, one has simply
\begin{equation}
\phi'(x')=\phi(x)\ ,
\end{equation}
and more generally for a collection of real or complex scalar fields,
a similar relation holds component by component. Note that at the quantum
level for the quantum field operators, such transformations are generated
by a representation $U(a,\Lambda)$ of the Poincar\'e group acting through
the adjoint action on the field operator in the Heisenberg picture,
\begin{equation}
\hat{\phi}(x')=U(a,\Lambda)\hat{\phi}(x)U^\dagger(a,\Lambda)\ .
\end{equation}

The Poincar\'e group being an abstract Lie group, possesses a collection 
of generators each of which is associated to each independent type of 
transformation. Thus for spacetime translations with the parameters 
$a^\mu$ one has the generators $P^\mu$, while for spacetime Lorentz 
(pseudo)rotations with the parameters ${\Lambda^\mu}_\nu$ one has the 
generators $M^{\mu\nu}=-M^{\nu\mu}$. At the abstract level, the corresponding
Lie algebra is given by the nonvanishing Lie brackets,
\begin{equation}
\begin{array}{r c l}
\left[P^\mu,P^\nu\right]&=&0\ \ \ ,\ \ \ 
\left[P^\mu,M^{\nu\rho}\right]=
i\left[\eta^{\mu\nu}P^\rho-\eta^{\mu\rho}P^\nu\right]\ ,\\
 & & \\
\left[M^{\mu\nu},M^{\rho\sigma}\right]&=&
-i\left[\eta^{\mu\rho}M^{\nu\sigma}-\eta^{\mu\sigma}M^{\nu\rho}-
\eta^{\nu\rho}M^{\mu\sigma}+\eta^{\nu\sigma}M^{\mu\rho}\right]\ ,
\end{array}
\end{equation}
where the last set of brackets determines the Lorentz
algebra. These ge\-ne\-ra\-tors induce finite Poincar\'e transformations
through their exponentiated action in the abstract realisation of the group,
\begin{equation}
U(a,\Lambda)=e^{ia_\mu P^\mu+\frac{1}{2}i\omega_{\mu\nu}M^{\mu\nu}}\ .
\end{equation}

According to the Noether theorem, any dynamics of which the Lagrange
function is invariant (possibly up to a total divergence)
under Poincar\'e transformations possesses
conserved quantities --- the Noether charges --- for solutions to the
classical equations of motion, which, in the
Hamiltonian formulation, generate through Poisson brackets these
symmetry transformations on phase space, and possess among themselves
Poisson brackets which coincide with the above Lie algebra 
brackets.\footnote{The notion of a dynamics invariant under a set of symmetry
transformations requires in fact that the action of the system, rather
than its Lagrange function, be invariant up to a surface term, since the
latter does not affect the equations of motion. If indeed the Lagrange function
is invariant only up to a surface term, central extensions of the symmetry
Lie brackets are also possible, already at the classical 
level.\cite{Jose-Luis} Nonetheless, Noether's theorem then remains valid,
though with a contribution of the induced surface terms to the conserved
charges.\cite{GovBook}} Hence, the Noether charges provide the explicit 
realisation within a given system of the associated symmetry generators 
in terms of the relevant degrees of freedom.

Thus for instance, in the case of the relativistic quantum scalar 
particle in the configuration space wave function 
representation,\cite{GovCOPRO2} the Poincar\'e algebra is realised by
the operators
\begin{equation}
P_\mu=-i\hbar\frac{\partial}{\partial x^\mu}=-i\hbar\partial_\mu\ \ \ ,\ \ \ 
M_{\mu\nu}=P_\mu x_\nu-P_\nu x_\mu\ ,
\end{equation}
where in the last expression the Lorentz covariant extension of the
usual orbital angular-momentum definition is recognised.

Likewise given any relativistic invariant local field theory,
Noether's theorem guarantees the existence of conserved charges
given by explicit functionals of the fields which generate
Poincar\'e transformations through Poisson brackets at the classical
level, and through commutation relations at the quantum level.
However, due to Lorentz covariance and spacetime locality
of the Lagrange function given as a space integral of a Lagrangian
density, it follows that the conservation condition is expressed\cite{GovBook}
through a divergenceless condition on a conserved current density,
of which the conserved charge is given by the integral over space
of its time component. In general terms,
\begin{equation}
\partial_\mu J^\mu=0\ \ \ ({\rm on\!-\!shell})\ \ \ ,\ \ \ 
Q=\int_{(\infty)}d^3\vec{x}\,J^{\mu=0}\ ,
\end{equation}
where $J^\mu$ denotes the Noether current density and $Q$ the associated
Noether charge, while ``on-shell'' stands for the fact that the
conservation property holds only for solutions to the classical
equations of motion.

In the case of a single real scalar field, the detailed analysis
of the Noether identities\cite{GovBook,QFT} associated to the 
invariance of the Lagrangian density under Poincar\'e transformations 
establishes that the Noether density is given by
\begin{equation}
{\mathcal T}_{\mu\nu}=\partial_\mu\phi\partial_\nu\phi-\eta_{\mu\nu}
{\mathcal L}\ ,
\end{equation}
a quantity which defines the energy-momentum density of the system.
In particular, the total energy-momentum content $P^\mu$ of the field is then
given as
\begin{equation}
P^\mu=\int_{(\infty)}d^3\vec{x}\,{\mathcal T}^{0\mu}\ :\ \ 
P^0=H_0=\int_{(\infty)}d^3\vec{x}\,{\mathcal H}_0\ \ ,\ \
\vec{P}=-\int_{(\infty)}d^3\vec{x}\,\pi_\phi\vec{\nabla}\phi\ ,
\end{equation}
where ${\mathcal H}_0$ stands for the canonical Hamiltonian density
of the system, and $\pi_\phi=\partial_0\phi$ for the momentum conjugate 
to the scalar field. For the total angular-momentum content, one also has
\begin{equation}
M^{\mu\nu}=\int_{(\infty)}d^3\vec{x}\,
\left[{\mathcal T}^{0\mu}x^\nu\,-\,{\mathcal T}^{0\nu}x^\mu\right]\ .
\end{equation}

Once given such expressions as well as the mode expansions of the
scalar field and its conjugate momentum in terms of the creation and
annihilation operators of its quanta, it is possible to also
determine the representations of these Poincar\'e charges in terms 
of the particle content of the field, whether at the classical
or the quantum level, as operators acting on Hilbert space. Thus,
once a normal ordering prescription is applied onto composite
operators --- whereby creation operators are always brought to the
left of annihilation operators\cite{GovCOPRO2,QFT} ---, one finds for
the energy-momentum content of the field,
\begin{equation}
\hat{P}^\mu=\int\frac{d^3\vec{k}}{(2\pi)^32\omega(\vec{k}\,)}\,
k^\mu\,a^\dagger(\vec{k}\,)a(\vec{k}\,)\ ,
\end{equation}
while its angular-momentum $\hat{M}^{\mu\nu}$ decomposes 
according to,\cite{QFT}
\begin{equation}
\begin{array}{r c l}
\hat{M}_{0j}&=&i\int\frac{d^3\vec{k}}{(2\pi)^32\omega(\vec{k}\,)}\,
a^\dagger(\vec{k}\,)\left[\omega(\vec{k}\,)\frac{\partial}{\partial k^j}\right]
a(\vec{k}\,)\ ,\\ 
 & & \\
\hat{M}_{j\ell}&=&
i\int\frac{d^3\vec{k}}{(2\pi)^32\omega(\vec{k}\,)}\,
a^\dagger(\vec{k}\,)\left[k_j\frac{\partial}{\partial k^{\ell}}\ -\
k_{\ell}\frac{\partial}{\partial k^j}\right] a(\vec{k}\,)\ .
\end{array}
\end{equation}
The expectation values of these quantities may thus be determined
for whatever quantum state the quantum field finds itself it.
In particular, 1-particle states define specific eigenstates of these
Poincar\'e generators (see below).

As is well known, representations of the Poincar\'e algebra 
ISO(1,3) are characterised by the eigenstates of its two
Casimir operators, namely the invariant energy 
$\hat{P}^2=\hat{P}^\mu\hat{P}^\nu\eta_{\mu\nu}$ which measures the 
invariant mass of field configurations, and the relativistic
invariant $\hat{W}^2$ of the Pauli-Lubanski 4-vector
$\hat{W}^\mu=\frac{1}{2}\epsilon^{\mu\nu\rho\sigma}
\hat{P}_{\nu}\hat{M}_{\rho\sigma}$,\footnote{Note that for a massive
particle in its rest-frame, the Pauli-Lubanski 4-vector does indeed reduce
to its total angular-momentum, {\it i.e.\/}, its spin.} which commutes 
with $\hat{P}^\mu$.

A massive representation of the Poincar\'e group is thus characterised
by the eigenvalues $\hat{P}^2=m^2$ and $\hat{W}^2=-m^2s(s+1)$, where
$m>0$ stands for its mass and $s$ for its spin, an integer or half-integer
valued quantity defining an irreducible representation of the group
SU(2), the universal covering group the 3-dimensional rotation group
SO(3) sharing the same Lie algebra of infinitesimal rotations in space.
For a massless representation, one has $\hat{P}^2=0$ and
$\hat{W}^2=0$. Such representations are characterised by the helicity
$s$ of the state, namely a specific representation of the helicity
group SO(2), the rotation subgroup of the Wigner little group for
a massless particle.\footnote{By definition, the Wigner little group
of a particle is the subgroup of the full Lorentz group leaving
invariant the particle's energy-momentum 4-vector. For a
massive particle, by going to its rest-frame, it is immediate
to establish that its little group is isomorphic to the space
rotation group SO(3) (at least for its component connected to the
identity transformation, namely homotopic to the identity) or
SO(D-1) in a $D$-dimensional Minkowski spacetime. For
a massless particle whose energy-momentum 4-vector is light-like,
a detailed analysis, based on the Lorentz algebra, 
shows that the little group is isomorphic to
the euclidean group E(D-2) for a Minkowski spacetime of dimension $D$
which combines the rotations SO(D-2) in the space directions
transverse to the particle momentum with specific combinations of
Lorentz boosts in the momentum direction with space rotations
around that momentum direction. At the quantum level, the notion of
spin is attached to massive particles, and determines a representation
of the SO(D-1) little group, while the notion of helicity is attached
to massless particles and is characterised by a representation
of the rotation subgroup SO(D-2) of the E(D-2) little group.
In four dimensions, $D=4$, both spin and helicity are thus specified
by a single integer of half-integer $s$.\cite{GovMexico}} For instance, 
for a light-like energy-momentum 4-vector $P^\mu=E(1,0,0,1)$, 
one has $W^\mu=M_{12}P^\mu$, so that $M_{12}$ takes the possible 
eigenvalues $\pm s$.

In the case of the scalar field, it is then straightforward to identify
the particle content of its Hilbert space. A 1-particle state 
$|\vec{k}\rangle=a^\dagger(\vec{k}\,)|0\rangle$ is characterised
by the eigenvalues
\begin{equation}
\hat{P}^0|\vec{k}\rangle=\omega(\vec{k}\,)|\vec{k}\rangle\ \ ,\ \ 
\hat{\vec{P}}|\vec{k}\rangle=\vec{k}\,|\vec{k}\rangle\ \ ,\ \ 
\hat{W}^2|\vec{k}\rangle=0\ ,
\end{equation}
thus showing that indeed, the quanta of such a quantum field may
be identified with particles of definite energy-momentum and mass
$m$, carrying a vanishing spin (in the massive case) or helicity
(in the massless case). Relativistic quantum field theories are thus
the natural framework in which to describe all the relativistic
quantum properties, including the processes of their annihilation
and creation in interactions, of relativistic quantum point-particles.
It is the Poincar\'e invariance properties, namely the relativistic
covariance of such systems, that also justifies, on account of
Noether's theorem, this physical interpretation.

One has to learn how to extend the above description to more
general field theories whose quanta are particles of nonvanishing
spin or helicity. Clearly, one then has to consider collections
of fields whose components also mix under Lorentz transformations,
namely nontrivial representations\footnote{A scalar field being invariant
under Lorentz transformations, is associated to the trivial
representation of the Lorentz group.} of the Lorentz group.

\section{Spinor Representations of the Lorentz Group\\
and Spin 1/2 Particles}
\label{Sec3}

\subsection{The Lorentz Group and Its Covering Algebra}
\label{Sec3.1}

Let us now consider the possibility that a collection of fields
$\phi_\alpha(x)$ (whether real or complex), distinguished by a
component index $\alpha$, provide a linear representation space
of the Poincar\'e group, whose action is defined according to\footnote{Note
well that the fields are taken to transform under the trivial representation
of the spacetime translation subgroup of the full Poincar\'e group. Hence
it is only for Lorentz transformations that we need to understand the
representation theory to be discussed in the present section.}
\begin{equation}
\begin{array}{r l c l}
- &{\rm translations}\ :&&\ \ 
\phi'_\alpha(x'=x+a)=\phi_\alpha(x)\ ,\\
 & & & \\
- &{\rm Lorentz\ transformations}\ :&&\ \ 
\phi'_\alpha(x'=\Lambda\cdot x)={S_\alpha}^\beta(\Lambda)\,\phi_\beta(x)\ .
\end{array}
\end{equation}
The sought for collection of fields is to provide a
representation space of the associated Lorentz $so(1,3)$ algebra,
\begin{equation}
\left[M^{\mu\nu},M^{\rho\sigma}\right]=
-i\left[\eta^{\mu\rho}M^{\nu\sigma}-
\eta^{\mu\sigma}M^{\nu\rho}-
\eta^{\nu\rho}M^{\mu\sigma}+
\eta^{\nu\sigma}M^{\mu\rho}\right]\ ,
\end{equation}
where the Lorentz boost generators $M^{0i}$ ($i=1,2,3$) must be
taken to be anti-hermitian, and the generators of rotations in space
$M^{ij}$ hermitian,\footnote{A finite dimensional representation of
a noncompact Lie algebra as is that of the Lorentz group is necessarily
nonunitary.}
\begin{equation}
\left(M^{0i}\right)^\dagger=-M^{0i}\ \ \ ,\ \ \ 
\left(M^{ij}\right)^\dagger=M^{ij}\ \ \ ,i,j=1,2,3\ .
\end{equation}

In order to exploit now a feature unique to 4-dimensional Minkowski
spacetime, let us introduce the following change of basis in the
complexified Lorentz Lie algebra,
\begin{equation}
L^i_\pm=\frac{1}{2}\left[L^i\pm iK^i\right]\ \ ,\ \ 
K^i=M^{0i}\ \ ,\ \ L^i=\frac{1}{2}\epsilon^{ijk}M^{jk}\ ,
\end{equation}
\begin{equation}
{L^i_\pm}^\dagger=L^i_\pm\ \ ,\ \ 
{K^i}^\dagger=-K^i\ \ ,\ \ 
{L^i}^\dagger=L^i\ .
\end{equation}
Note that the generators $L^i_\pm$ combine a Lorentz boost
in the direction $i$ with a rotation around that direction
in opposite directions, hence in effect defining chiral rotations
in spacetime, and leading to hermitian generators for the complexified
Lorentz algebra. A direct calculation then readily finds that in 
terms of these chiral generators $L^i_\pm$, the Lorentz algebra 
factorises into the direct sum of two $su(2)_\pm$ algebras,
\begin{equation}
\left[L^i_\pm,L^j_\pm\right]=i\epsilon^{ijk}L^k_\pm\ \ \ ,\ \ \ 
\left[L^i_\pm,L^j_\mp\right]=0\ .
\end{equation}
In other words, the complexification of the $so(1,3)$ Lorentz
algebra is isomorphic to the algebra 
$su(2)_+\oplus su(2)_-=sl(2,\mathbb C)$.\footnote{The relation to
SL(2,$\mathbb C$) is discussed hereafter.} Consequently, the
universal covering algebra (over $\mathbb C$) of the Lorentz group 
algebra $so(1,3)$ is that of the group SU(2)$_+\times$SU(2)$_-$.

The obvious advantage of this result is that the representation
theory of the Lorentz group in 4-dimensional Minkowski spacetime
may be understood in terms of representations of the SU(2) group,
which are well known from the notion of spin and angular-momentum
in nonrelativistic quantum mechanics. To each of the factors 
$su(2)_\pm$ one must associate an integer or half-integer value 
$j_\pm$ which determines a specific irreducible representation of SU(2), 
namely that of ``spin'' $j_\pm$. Thus finite dimensional irreducible 
representations of the Lorentz group SO(1,3) are characterised by 
a pair of integer or half-integer values $(j_+,j_-)$. The trivial 
representation is that characterised by $(j_+,j_-)=(0,0)$. Next one has 
the two inequivalent representations $(j_+,j_-)=(1/2,0)$ and 
$(j_+,j_-)=(0,1/2)$, which will be seen to play a fundamental role 
hereafter. One may also have for instance 
$(j_+,j_-)=(1/2,1/2),(1,0),(0,1)$, etc. In fact, since
in SU(2), all representations may be obtained through tensor products
of the fundamental $j=1/2$ spinor representation, likewise for the
Lorentz group, all its finite dimensional irreducible representations
may be obtained through tensor products of the two inequivalent
spinor representations $(j_+,j_-)=(1/2,0)$ and $(j_+,j_-)=(0,1/2)$, 
which are thus the two fundamental representations of the Lorentz group, 
known as the Weyl spinors of opposite right- or left-handed chiralities,
respectively.

Given any such $(j_+,j_-)$ Lorentz representation, its spin content
may also easily be identified. Indeed, in terms of the chiral
generators $L^i_\pm$, the SO(3) angular-momentum generators $L^i$
are obtained simply as the direct sum $L^i=L^i_++L^i_-$. Thus, 
the spin content of a given $(j_+,j_-)$ representation is simply 
obtained through the usual rules for spin reduction of tensor products 
of SU(2) representations. Consequently, a $(j_+,j_-)$ representations contains 
spin representations of $so(3)=su(2)$ of values spanning the range
\begin{equation}
|j_+-j_-|,\ |j_+-j_-|+1,\ \cdots,\ j_++j_-\ .
\end{equation}

Finally, a given $(j_+,j_-)$ Lorentz representation is not invariant
under parity. Indeed, under this transformation in space, the Lorentz
boost ge\-ne\-ra\-tors $K^i$ change sign whereas the angular-momentum
ones $L^i$ do not. Hence under parity, the two classes of chiral
operators $L^i_\pm$ are simply exchanged, inducing the correspondence
under parity of the representations $(j_+,j_-)$ and $(j_-,j_+)$.
Consequently, when the Lorentz group SO(1,3) is extended to also
include the parity transformation, its irreducible representations
are to be combined into the direct sums $(j_+,j_-)\oplus(j_-,j_+)$ in the
case of distinct values for $j_+$ and $j_-$.

\newpage

Given all these considerations, one may list the representations
which are invariant under parity and correspond to the lowest
spin or helicity content possible,
\begin{equation}
\begin{array}{r c l}
(0,0)\ &:&\ \ \ {\rm scalar\ field}\ \phi;\\
(1/2,0)\oplus(0,1/2)\ &:&\ \ {\rm Dirac\ spinor}\ \psi;\\
(1/2,1/2)\ &:&\ \ {\rm vector\ field}\ A_\mu;\\
(1,0)\oplus(0,1)\ &:&\ \ {\rm electromagnetic\ field\ strength\ tensor}\\
& & \ \ 
F_{\mu\nu}=\partial_\mu A_\nu-\partial_\nu A_\mu\simeq(\vec{E},\vec{B})\ .
\end{array}
\end{equation}

The simplest $\mathcal N=1$ supersymmetry realisation in 4-dimensional
Minkowski spacetime in fact relates scalar and spinor fields, as well
as spinor and vector fields. The fundamental Lorentz spinors
correspond to the right- and left-handed
Weyl spinors $(1/2,0)$ and $(0,1/2)$, respectively, which are
exchanged under the parity transformation. In terms of the quanta
of such fields, Weyl spinors describe massless
particles of fixed helicity $s=\pm 1/2$ equal to the chirality
$\pm 1/2$ of the Weyl spinor, and antiparticles of the opposite
helicity $s=\mp 1/2$. Weyl spinors must be combined in
order to describe massive spin 1/2 particles, one possibility being
the celebrated Dirac spinor and its Dirac equation, describing massive
spin 1/2 particles and antiparticles invariant under parity, 
which is to be discussed hereafter.

\subsection{An Interlude on SU(2) Representations}
\label{Sec3.2}

Let us pause for a moment to recall a few well known facts
concerning SU(2) representations, that will become relevant
in the next section. The $su(2)$ Lie algebra is spanned by three
generators $T^i$ $(i=1,2,3)$ with the Lie bracket algebra
\begin{equation}
\left[T^i,T^j\right]=i\epsilon^{ijk}\,T^k\ \ ,\ \ 
\epsilon^{123}=+1\ .
\end{equation}

As is the case for any SU(N) algebra, {\it a priori\/}, SU(2) possesses
two fundamental representations of dimension two, complex conjugates 
of one another, namely the spinor representations of SU(2) or SO(3).
There is the ``covariant'' 2-dimensional representation $\underline{\bf 2}$, 
a vector space spanned by covariant complex valued doublet vectors 
$a_\alpha$ $(\alpha=1,2)$ transforming under a SU(2) group element 
${U_\alpha}^\beta$, with $U^\dagger=U^{-1}$ and ${\rm det}\,U=1$, as
\begin{equation}
{a'}_\alpha={U_\alpha}^\beta\,a_\beta\ .
\end{equation}
This representation is also associated to the generators
\begin{equation}
T^i=\frac{1}{2}\sigma_i\ \ \ ,\ \ \ i=1,2,3\ ,
\end{equation}
the $\sigma_i$ being the usual Pauli matrices,\footnote{The position of
the index $i$ is important in these relations, for reasons to become clear
later on.}
\begin{equation}
\sigma_1=\left(\begin{array}{c c}
		0 & 1 \\
		1 & 0 
		\end{array}\right)\ \ ,\ \ 
\sigma_2=\left(\begin{array}{c c}
		0 & -i \\
		i & 0 
		\end{array}\right)\ \ ,\ \ 
\sigma_3=\left(\begin{array}{c c}
		1 & 0 \\
		0 & -1 
		\end{array}\right)\ .
\end{equation}

Correspondingly, the ``contravariant'' complex conjugate 2-dimensional
representation $\overline{\underline{\bf 2}}$,
spanned by vectors $a^\alpha$ ($\alpha=1,2)$, consists of complex valued
vectors transforming under SU(2) group elements as
\begin{equation}
{a'}^\alpha=a^\beta\,{U^\dagger_\beta}^\alpha=
a^\beta\,{U^{-1}_\beta}^\alpha\ ,
\end{equation}
and associated to the generators $T^i=\sigma^*_i/2$.

Similar considerations apply to the SU(N) case. The fact that in this general
case these are the two fundamental representations is related to the existence
of two SU(N)-invariant tensors, namely the Kronecker symbols
${\delta^\alpha}_\beta$ and ${\delta_\alpha}^\beta$ and 
the totally antisymmetric symbols $\epsilon^{\alpha_1\cdots\alpha_N}$ 
and $\epsilon_{\alpha_1\cdots\alpha_N}$,
which themselves are directly connected to the defining properties of
SU(N) matrices, namely the fact that they are unitary, $U^\dagger=U^{-1}$,
and of unit determinant, ${\rm det}\,U=1$. Particularised to the SU(2) case,
these simple properties may easily be checked. Indeed, using the transformation
rules recalled above for co- and contra-variant indices under the SU(2)
action, one has, for instance,
\begin{equation}{\delta'_\alpha}^\beta={U_\alpha}^{\alpha_1}\,
{\delta_{\alpha_1}}^{\beta_1}\,{U^\dagger_{\beta_1}}^\beta=
{\delta_\alpha}^\beta\ ,
\end{equation}
a result which readily follows from the unitarity property of SU(2) elements,
$U^\dagger=U^{-1}$. Likewise for the $\epsilon_{\alpha\beta}$ tensor,
for instance,
\begin{equation}\epsilon'_{\alpha\beta}=
{U_\alpha}^{\alpha_1}\,{U_\beta}^{\beta_1}\,\epsilon_{{\alpha_1}{\beta_1}}=
\epsilon_{\alpha\beta}\ ,
\end{equation}
a result which follows from the unit determinant value, ${\rm det}\,U=1$.

In the general SU(N) case, these considerations imply that the
$N$-dimensional contravariant representation $\overline{\underline{\bf N}}$,
the complex conjugate of the $N$-dimensional covariant one $\underline{\bf N}$,
is also equivalent to the totally antisymmetry representation obtained
through the $(N-1)$-times totally antisymmetrised
tensor product of the latter representation with itself,
\begin{equation}
a_{\alpha_1\cdots\alpha_{N-1}}=\epsilon_{\alpha_1\cdots\alpha_{N-1}\beta}
a^\beta\ .
\end{equation}
However, the SU(2) case is distinguished in this regard by the fact
that this transformation also defines a unitary transformation
on representation space. In other words, the relations
\begin{equation}
a^\alpha=\epsilon^{\alpha\beta}\,a_\beta\ \ \ ,\ \ \ 
a_\alpha=\epsilon_{\alpha\beta}\,a^\beta\ ,
\end{equation}
establish the unitary equivalence of the two 2-dimensional SU(2)
representations $\underline{\bf 2}$ and $\overline{\underline{\bf 2}}$.
For example, one may check that these quantities do indeed transform
under SU(2) according to the rules associated to the position of
the index $\alpha$, using the invariant properties of the two available
SU(2) invariant tensors, for instance,
\begin{equation}
{a'}^\alpha=\epsilon^{\alpha\beta}{U_\beta}^\gamma\,a_\gamma=
{\left(U^{-1}\right)_\beta}^\alpha\,\epsilon^{\beta\gamma}\,a_\gamma=
a^\beta{U^{-1}_\beta}^\alpha=a^\beta\,{U^\dagger_\beta}^\alpha\ .
\end{equation}

The unitary equivalence between the two 2-dimensional SU(2)
representations is thus determined by the unitary matrix
\begin{equation}
\epsilon^{\alpha\beta}=\left(i\sigma_2\right)^{\alpha\beta}\ \ ,\ \ 
\epsilon_{\alpha\beta}=\left(-i\sigma_2\right)_{\alpha\beta}\ .
\end{equation}
This matrix being also antisymmetric, means that in fact the
2-dimensional SU(2) representation ($\underline{\bf 2}$ or 
$\overline{\underline{\bf 2}}$) is a pseudoreal representation.
Contrary to SU(N) with $N>2$ for which the $\underline{\bf N}$ and
$\overline{\underline{\bf N}}$ representations are the two
inequivalent fundamental complex representations, in the SU(2) case
there is only a single fundamental representation which is also
pseudoreal. This is the SU(2) spinor representation.
Consequently, it is also clear that all higher spin SU(2) representations
are either real or pseudoreal, namely are unitarily equivalent to
their complex conjugate representations with a unitary matrix
defining this equivalence which is either symmetric or antisymmetric,
respectively, according to whether they are obtained with an even
or an odd number of tensor product factors of the fundamental spinor
representation. In other words, all integer spin SU(2) representations
are strictly real, whereas all half-integer spinor representations are
pseudoreal representations. In fact, all integer spin representations
are actual representations of SO(3)=SU(2)/${\mathbb Z}_2$ corresponding
to all tensor representations of arbitrary rank, whereas all half-integer
spin representations are representations of SU(2), the universal
covering group of SO(3), but not of SO(3) itself because of the ${\mathbb Z}_2$
factor, the center of SU(2) (a rotation of angle $2\pi$ in a half-integer
spin or spinor representation is given by $(-\one)$, but by $\one$ in an
integer spin or tensor representation). This distinction between tensor and
spinor representations of the rotation group SO(3) is related to
the fact that SO(3) is a doubly-connected Lie group, a 3-dimensional
manifold equivalent to the solid 2-sphere of radius $\pi$ and
with opposite points identified, obtained as the quotient of
SU(2) by its center ${\mathbb Z}_2$ taking on the value $(-1)$ 
(resp. $(+1)$) for any rotation by $2\pi$ (resp. $4\pi$) around any 
given axis. In contradistinction the SU(2) manifold is that of the 
3-sphere,\footnote{This is readily established by considering a
real parametrisation of $2\times 2$ complex matrices, and imposing
the constraints of unitarity and unit determinant defining the
SU(2) matrix group.} which is simply connected.

This detailed characterisation of SU(2) representations
enables the direct construction of quantities which are SU(2)
invariants. For instance, consider two covariant spinors $a_\alpha$
and $b_\beta$. Since the tensor product of the spin 1/2 representation
with itself includes the trivial representation of zero spin,
the associated SU(2) invariant must exist, and is given by the
explicit SU(2) invariant contraction of the different indices
in a manner involving the two invariant tensors available,
\begin{equation}
\epsilon^{\alpha\beta}a_\alpha b_\beta=a_\alpha b^\beta=
a_\alpha\,{\delta^\alpha}_\beta\,b^\beta\ ,
\end{equation}
showing how the singlet component may be identified within the
tensor products $\underline{\bf 2}\otimes\underline{\bf 2}$
or $\underline{\bf 2}\otimes\overline{\underline{\bf 2}}$. This simple
rule for the construction of SU(2) invariants for SU(2) tensor
products will thus readily extend to the construction of
Lorentz invariant quantities, since the SO(1,3) Lorentz group
shares the same algebra as the SU(2)$_+\times$SU(2)$_-$ group,
of which the two independent spinor representations define
the two fundamental Weyl spinor representations of the Lorentz
group.

\subsection{The Fundamental Lorentz Representations:\\
Weyl Spinors}
\label{Sec3.3}

We have seen how, on the basis of the chiral SU(2)$_+\times$SU(2)$_-$
group, it is possible to readily identify the finite dimensional representation
theory of the Lorentz group SO(1,3). Let us now discuss yet another
construction of its two fundamental Weyl spinor representations, which
is also of importance in the construction of supersymmetric field
theories. The present discussion shall also make explicit why
the universal covering group of the Lorentz group SO(1,3) is
the group SL(2,$\mathbb C$) of complex 2$\times$2 matrices of unit
determinant. We shall thus establish the relation, at the level of the
corresponding Lie algebras,
\begin{equation}
so(1,3)_{\mathbb C}=su(2)_+\oplus su(2)_-=sl(2,{\mathbb C})\ .
\end{equation}

Let us introduce the notation
\begin{equation}
\sigma_\mu=(\one,\sigma_i)\ \ \ ,\ \ \ 
\sigma^\mu=(\one,\sigma^i)=(\one,-\sigma_i)\ ,
\end{equation}
where the space index $i$ carried by the usual Pauli matrices
is raised and lowered according to our choice of signature for the Minkowski
spacetime metric, namely $\eta_{\mu\nu}={\rm diag}\,(+---)$.
Consider now an arbitrary spacetime 4-vector $x^\mu$, and construct
the 2$\times$2 hermitian matrix
\begin{equation}
X=x^\mu\sigma_\mu=\left(\begin{array}{c c}
		x^0+x^3 & x^1-ix^2 \\
		x^1+ix^2 & x^0-x^3
			\end{array}\right)\ .
\end{equation}
Note that conversely, any 2$\times$2 hermitian matrix $X=X^\dagger$
possesses such a decomposition, and may thus be associated to some
spacetime 4-vector $x^\mu$ through the above relation. In particular,
the determinant of any such matrix is equal to the Lorentz invariant
inner product of the associated 4-vector with itself,
\begin{equation}
{\rm det}\,X=x^2=\eta_{\mu\nu}x^\mu x^\nu\ .
\end{equation}

Consider now an arbitrary SL(2,$\mathbb Z$) group element $M$, 
thus of unit determinant, ${\rm det}\,M=1$, and its adjoint action 
on any hermitian matrix $X$ as
\begin{equation}
X'=M\,X\,M^\dagger\ .
\end{equation}
It should be clear that the transformed matrix itself is hermitian,
${X'}^\dagger=X'$, hence possesses a decomposition in terms of
a 4-vector ${x'}^\mu$, $X'={x'}^\mu\sigma_\mu$, of which the Lorentz
invariant takes the value
\begin{equation}
{x'}^2={\rm det}\,X'={\rm det}\,MXM^\dagger={\rm det}\,X=x^2\ .
\end{equation}
In other words, any SL(2,$\mathbb C$) transformation induces a
Lorentz transformation on the 4-vector $x^\mu$. The group SL(2,$\mathbb C$)
determines a covering group of the Lorentz group SO(1,3). In fact,
it is the universal covering of the latter, as the discussion hereafter
in terms of its fundamental representations establishes. This conclusion
is thus analogous to that which states that SU(2) is the universal
covering group of the group SO(3) of spatial rotations. Indeed, the above
discussion may also be developed in the latter case, simply by
ignoring the time component of the matrices $\sigma_\mu$ and
then restricting further the matrices $X$ to be both hermitian
and traceless.

This conclusion having been reached, the next question is: how
does one construct the two fundamental Weyl spinor representations
of SO(1,3) in terms of SL(2,$\mathbb C$) representations? An arbitrary
SL(2,$\mathbb C$) matrix $M$ with ${\rm det}\,M=1$
may be decomposed according to
\begin{equation}
M=e^{(a_j+ib_j)\sigma_j}\ \ \ ,\ \ \ 
M^\dagger=e^{(a_j-ib_j)\sigma_j}\ ,
\end{equation}
where $a_j$ and $b_j$ ($j=1,2,3$) are triplets of real numbers.
In these terms, the SU(2)$_+\times$SU(2)$_-$ structure of the $sl(2,\mathbb C)$
algebra should again be obvious, with in particular the hermitian (related
to space rotations) and antihermitian (related to Lorentz boosts)
components of the Lie algebra.\footnote{Note that the counting of independent
parameters of the different SO(1,3), SU(2)$_+\times$SU(2)$_-$ and
SL(2,$\mathbb C$) groups also matches this correspondence. These three Lie
groups are all 6-dimensional.} In the case of SU(2), the additional
property is that the matrices defining the group are also unitary,
$U^\dagger=U^{-1}$. As a consequence, we have seen that the
two fundamental 2-dimensional SU(2) representations, complex conjugates of
one another, are unitarily equivalent. In the SL(2,$\mathbb C$) case,
these two representations are no longer equivalent. However, because
of the property ${\rm det}\,M=1$, $\epsilon^{\alpha\beta}$
and $\epsilon_{\alpha\beta}$ still define SL(2,$\mathbb C$) invariant
tensors, that may be used to raise and lower indices. Because of
this latter fact, there exist only two independent fundamental
2-dimensional representations of SL(2,$\mathbb C$), in direct
correspondence with the two chiral Weyl spinors considered
previously.

Another way of arguing the same conclusion is as follows. Given
a matrix $M\in$\ SL(2,$\mathbb C$), each of the matrices $M$,
$M^{-1}$, $M^*$ and $(M^*)^{-1}$ defines {\it a priori\/} another
2-dimensional representation of the same group. As pointed out above, 
$M$ and $M^*$ are necessarily not unitarily equivalent. However, 
$M$ and $M^{-1}$ on the one hand, and $M^*$ and $(M^*)^{-1}$ on 
the other hand, are each unitarily equivalent in pairs, using the invariant 
tensors $\epsilon^{\alpha\beta}$ and $\epsilon_{\alpha\beta}$ because
of the property ${\rm det}\,M=1$ for these $2\times 2$ matrices.

In conclusion, first we have the right-handed Weyl spinor
representation $(1/2,0)$, $\psi_\alpha$ or $\psi^\alpha$, such that
\begin{equation}
\psi^\alpha=\epsilon^{\alpha\beta}\,\psi_\beta\ \ ,\ \ 
\psi_\alpha=\epsilon_{\alpha\beta}\,\psi^\beta\ \ ,\ \ 
\epsilon^{12}=+1\ \ ,\ \ 
\epsilon_{12}=-1\ ,
\end{equation}
and transforming under SL(2,$\mathbb C$) according to
\begin{equation}
{\psi'}_\alpha={M_\alpha}^\beta\,\psi_\beta\ \ ,\ \ 
{\psi'}^\alpha=\psi^\beta\,{\left(M^{-1}\right)_\beta}^\alpha\ .
\end{equation}
Likewise, the left-handed Weyl spinor representation $(0,1/2)$,
$\bar{\psi}_{\dot{\alpha}}$ or $\bar{\psi}^{\dot{\alpha}}$, is such that
\begin{equation}
\overline{\psi}^{\dot{\alpha}}=\epsilon^{\dot{\alpha}\dot{\beta}}\,
\overline{\psi}_{\dot{\beta}}\ \ ,\ \ 
\overline{\psi}_{\dot{\alpha}}=\epsilon_{\dot{\alpha}\dot{\beta}}\,
\overline{\psi}^{\dot{\beta}}\ \ ,\ \ 
\epsilon^{\dot{1}\dot{2}}=+1\ \ ,\ \ 
\epsilon_{\dot{1}\dot{2}}=-1\ ,
\end{equation}
each of these spinors transforming according to
\begin{equation}
{\overline{\psi}'}_{\dot{\alpha}}={M^*_{\dot{\alpha}}}^{\dot{\beta}}\,
\overline{\psi}_{\dot{\beta}}\ \ ,\ \ 
{\overline{\psi}'}^{\dot{\alpha}}=\overline{\psi}^{\dot{\beta}}\,
{\left((M^*)^{-1}\right)_{\dot{\beta}}}^{\dot{\alpha}} \ .
\end{equation}
Here, in order to distinguish these two SL(2,$\mathbb C$) representations,
or equivalently the two Weyl spinors, the van der Waerden dotted and
undotted index notation has been introduced. This notation proves
particularly valuable for the construction of manifestly supersymmetric
invariant Lagrangian densities.

The undotted indices $\alpha,\beta$, on the one hand, and 
dotted indices $\dot{\alpha},\dot{\beta}$, on the other hand,
have the same meaning as the $\alpha,\beta$ indices for the SU(2) 
spinor representations. Consequently, Lorentz invariant
quantities are readily constructed in terms of the Weyl spinors
$\psi^\alpha$ and $\overline{\psi}_{\dot{\alpha}}$, through simple contraction
of the indices using the invariant tensors available. Furthermore,
given that we have
\begin{equation}
{x'}^\mu\sigma_\mu=X'=MXM^\dagger=M\left(x^\mu\sigma_\mu\right)M^\dagger\ ,
\end{equation}
it follows that the SL(2,$\mathbb C$) or SO(1,3) Lorentz
transformation properties of the matrices $\sigma_\mu$ are those
characterised by the index structure,
\begin{equation}
\sigma^\mu\ :\ \ \ 
\left(\sigma^\mu\right)_{\alpha\dot{\alpha}}\ \ ,\ \ 
\sigma_\mu=(\one,\sigma_i)\ \ ,\ \ 
\sigma^\mu=(\one,-\sigma_i)=(\one,\sigma^i)\ .
\end{equation}
By raising the indices, one introduces the quantities
\begin{equation}
\overline{\sigma}^\mu\ :\ \ \ 
\left(\overline{\sigma}^\mu\right)^{\dot{\alpha}\alpha}=
\epsilon^{\dot{\alpha}\dot{\beta}}\,
\epsilon^{\alpha\beta}\,\left(\sigma_\mu\right)_{\beta\dot{\beta}}\ \ ,\ \ 
\overline{\sigma}^\mu=(\one,\sigma_i)\ \ ,\ \ 
\overline{\sigma}_\mu=(\one,-\sigma_i)\ .
\end{equation}
Note that these properties also justify why indeed a 4-vector $A_\mu$
is equivalent to the $(1/2,1/2)=(1/2,0)\oplus(0,1/2)$ Lorentz
representation, $A^\mu\sigma_{\mu\alpha\dot{\alpha}}=A_{\alpha\dot{\alpha}}$,
$A^\mu{\overline{\sigma}_{\mu}}^{\dot{\alpha}\alpha}=
\overline{A}^{\dot{\alpha}\alpha}$.

Let us now consider different Weyl spinors $\psi$, $\chi$,
$\overline{\psi}$, $\overline{\chi}$, ... and the Lorentz invariant
spinor bilinears that may constructed out of these quantities.
For this purpose, it is important to realise that such field degrees
of freedom, at the classical level, need to be described in terms of
Grassmann odd variables, namely variables $\theta_1$, $\theta_2$, $\cdots$
which anticommute with one another, $\theta_2\theta_1=-\theta_1\theta_2$,
in contradistinction to commuting variables used for
fields describing particles of integer spin and obeying Bose--Einstein
statistics. The reasons for this necessary choice will be discussed
somewhat further later on, but at this stage, it suffices to say that
spinorial fields are associated to particles of half-integer spin
which should thus obey Fermi--Dirac statistics with the consequent
Pauli exclusion principle, a result which is readily achieved provided
Grassmann odd degrees of freedom are used even at the classical level.
The associated Grassmann graded Poisson brackets\cite{GovBook} then 
correspond, at the quantum level, to anticommutation rather than 
commutation relations for the degrees of freedom, ensuring 
the Fermi--Dirac statistics. The anticommuting character of the Weyl 
spinors hereafter is an important fact to always keep in mind when 
performing explicit calculations.

Since dotted and undotted indices cannot be contracted with one
another in a Lorentz invariant way, there are only two types of
Lorentz invariant spinor bilinears that may be considered. By
definition, those associated to undotted spinors write as,
\begin{equation}
\begin{array}{r c l}
\psi\chi&=&\psi^\alpha\,\chi_\alpha=
\epsilon^{\alpha\beta}\,\psi_\beta\chi_\alpha=
-\epsilon^{\alpha\beta}\psi_\alpha\chi_\beta\\
 & & \\
&=&-\psi_\alpha\,\chi^\alpha=\chi^\alpha\,\psi_\alpha=\chi\,\psi\ .
\end{array}
\end{equation}
The convention here, implicit throughout the supersymmetry literature,
is that for undotted spinors, the Lorentz invariant
contraction denoted $\psi\,\chi$ without displaying the indices,
is that in which the undotted indices are contracted from
top-left to bottom-right. Note that the Grassmann odd property of
the Weyl spinors has been used to derive the above identity,
$\psi\chi=\chi\psi$.

In contradistinction for dotted spinors, the convention is that
the contraction is taken from bottom-left to top-right, namely
\begin{equation}
\begin{array}{r c l}
\overline{\psi}\,\overline{\chi}&=&
\overline{\psi}_{\dot{\alpha}}\,\overline{\chi}^{\dot{\alpha}}=
\epsilon_{\dot{\alpha}\dot{\beta}}\,\overline{\psi}^{\dot{\beta}}\,
\overline{\chi}^{\dot{\alpha}}
=-\epsilon_{\dot{\alpha}\dot{\beta}}\,\overline{\psi}^{\dot{\alpha}}\,
\overline{\chi}^{\dot{\beta}}\\
 & & \\
&=&-\overline{\psi}^{\dot{\alpha}}\,\overline{\chi}_{\dot{\alpha}}=
\overline{\chi}_{\dot{\alpha}}\,\overline{\psi}^{\dot{\alpha}}=
\overline{\chi}\,\overline{\psi}\ .
\end{array}
\end{equation}
Further identities that may be established in a likewise manner are,
\begin{equation}
\left(\psi\,\chi\right)^\dagger=
\overline{\chi}\,\overline{\psi}=
\overline{\psi}\,\overline{\chi}\ \ ,\ \
\left(\overline{\psi}\,\overline{\chi}\right)^\dagger=
\chi\,\psi=\psi\,\chi\ .
\end{equation}

For the construction of Lorentz covariant spinor bilinears, one
has to also involve the matrices $\sigma_\mu$ and $\overline{\sigma}_\mu$.
Thus for instance, we have the quantities transforming as 4-vectors
under Lorentz transformations,
\begin{equation}
\psi\,\sigma^\mu\,\overline{\chi}=
\psi^\alpha\,{\sigma^\mu}_{\alpha\dot{\beta}}\,\overline{\chi}^{\dot{\beta}}
\ \ ,\ \ 
\overline{\psi}\,\overline{\sigma}^\mu\,\chi=
\overline{\psi}_{\dot{\alpha}}\,\overline{\sigma}^{\mu\dot{\alpha}\beta}\,
\chi_\beta\ .
\end{equation}
Such quantities also obey a series of identities, for instance,
\begin{equation}
\chi\,\sigma^\mu\,\overline{\psi}=-
\overline{\psi}\,\overline{\sigma}^\mu\,\chi\ \ ,\ \ 
\chi\,\sigma^\mu\,\overline{\sigma}^\nu\,\psi=
\psi\,\sigma^\nu\,\overline{\sigma}^\mu\,\chi\ ,
\end{equation}
\begin{equation}
\left(\chi\,\sigma^\mu\,\overline{\psi}\right)^\dagger=
\psi\,\sigma^\mu\,\overline{\chi}\ \ ,\ \ 
\left(\chi\,\sigma^\mu\,\overline{\sigma}^\nu\,\psi\right)^\dagger=
\overline{\psi}\,\overline{\sigma}^\nu\,\sigma^\mu\,\overline{\chi}\ .
\end{equation}
Identities of this type enter the explicit construction
of supersymmetric invariant field theories.

\subsection{The Dirac Spinor}
\label{Sec3.4}

As mentioned earlier, Weyl spinors are not parity invariant
representations of the Lorentz group. The fundamental parity invariant
representation is obtained as the direct sum of a right- and a left-handed
Weyl spinor, leading to the Dirac spinor, a 4-dimensional spinor
representation of the Lorentz group, which is irreducible for the
Lorentz group SO(1,3) extended to also include the parity transformation.
Furthermore, since the dotted and undotted notation is not as familiar as the
Dirac spinor construction, the latter will now be considered in detail
through its relation to the previous discussion.

Given the 4-dimensional Dirac representation, it is useful to
combine the $\sigma^\mu$ and $\overline{\sigma}^\mu$ matrices into
a collection of 4$\times$4 matrices,
\begin{equation}
\gamma^\mu=\left(\begin{array}{c c}
	0 & \sigma^\mu \\
	\overline{\sigma}^\mu & 0 
		\end{array}\right)\ \ \ ,\ \ 
\gamma_5=i\gamma^0 \gamma^1 \gamma^2 \gamma^3=
\left(\begin{array}{c c}
	\one & 0 \\
	0 & -\one 
	\end{array}\right) \ ,
\end{equation}
known as the Dirac matrices. As a matter of fact, the above definition
provides a specific representation of the Dirac-Clifford algebra that
these matrices obey,
\begin{equation}
\begin{array}{r c l}
\left\{\gamma^\mu,\gamma^\nu\right\}=2\eta^{\mu\nu}\ \ \ &,&\ \ \ 
\left\{\gamma^\mu,\gamma_5\right\}=0\ ,\\
 & & \\
{\gamma^\mu}^\dagger=\gamma^0 \gamma^\mu \gamma^0\ \ \ &,&\ \ \ 
\gamma^\dagger_5=\gamma_5\ \ ,\ \ 
\gamma^2_5=\one\ .
\end{array}
\end{equation}
Other matrix representations of this algebra exist (among which that
originally constructed by Dirac himself\cite{QFT} when he discovered 
the celebrated Dirac equation). However in a Minkowski spacetime of 
even dimension, all these representations are unitarily equivalent.\cite{PvN}
The above representation of the Dirac-Clifford algebra is known as 
the chiral or Weyl representation, since the chiral projection operator 
$\gamma_5$ is then diagonal.

Being the direct sum of a right- and a left-handed Weyl spinor,
within the chiral representation a Dirac spinor decomposes according to
\begin{equation}
\psi^{\rm Dirac}_{(4)}=\left(\begin{array}{c}
	\psi_\alpha \\ \overline{\chi}^{\dot{\alpha}}
		\end{array}\right)\ ,
\end{equation}
where $\psi_\alpha$ is a $(1/2,0)$ right-handed Weyl spinor,
and $\overline{\chi}^{\dot{\alpha}}$ a $(0,1/2)$ left-handed one. These
two chiral components are indeed projected from the Dirac spinor
through the projectors
\begin{equation}
P_R=\frac{1}{2}\left(1+\gamma_5\right)\ \ ,\ \ 
P_L=\frac{1}{2}\left(1-\gamma_5\right)\ ,
\end{equation}
with the properties,
\begin{equation}
P^2_R=P_R\ \ ,\ \ P^2_L=P_L\ \ ,\ \ 
P_L P_R = 0 = P_R P_L\ .
\end{equation}
 
{\it A priori\/}, the two Weyl spinors $\psi_\alpha$ and
$\overline{\chi}^{\dot{\alpha}}$ are independent spinors,
leading to the construction of an actual Dirac spinor with
these many independent degrees of freedom. However, it could be
that these two Weyl spinors are complex conjugates of one another,
in which case the above construction defines what is known as
a Majorana spinor,
\begin{equation}
\psi^{\rm Majorana}_{(4)}=\left(\begin{array}{c}
	\psi_\alpha \\ \overline{\psi}^{\dot{\alpha}}
	\end{array}\right)\ .
\end{equation}
A Majorana spinor is to a Dirac spinor what a real scalar field
is to a complex scalar field. Namely, whereas the quanta of a
real scalar field are particles that cannot be distinguished
from their antiparticles (they do not carry a conserved quantum
number that could distinguish them, such as the electric charge), the
quanta of a complex scalar field are classified in terms of
particles and antiparticles, which may be distinguished according
to a conserved quantum number, for instance their electric charge,
associated to the global symmetry invariance under arbitrary
spacetime constant variations in the complex phase of the complex
scalar field.\cite{GovCOPRO2} Likewise for the above spinors, since
the Majorana spinor obeys some sort of restriction under complex
conjugation (its Weyl components of opposite chiralities are
related through complex conjugation), a Majorana spinor describes
spin or helicity 1/2 particles which are their own antiparticles,
and thus cannot carry a conserved quantum number such as the electric
charge.\footnote{Consequently, among quarks and leptons, only
neutrinos could possibly be Majorana particles. The experimental
verdict is still out, and is an important issue in the quest
for the fundamental unification of all interactions and particles.}
In contradistinction, the quanta associated to a Dirac
spinor may be distinguished in terms of particles and their
antiparticles carrying opposite values of a conserved quantum number,
such as for instance the electric charge (or baryon or lepton number),
associated to a symmetry under arbitrary global phase transformations 
of the Dirac spinor. As the above construction clearly shows,
in a 4-dimensional Minkowski spacetime, one cannot have both
a Weyl and a Majorana condition imposed on a Dirac spinor.
In such a case, one has either only Dirac spinors, Majorana spinors,
or Weyl spinors of definite chirality, while the fundamental
constructs of Lorentz covariant spinors are the two fundamental
right- and left-handed Weyl spinors. In fact, it may be shown,\cite{PvN}
using the properties of the Dirac-Clifford algebra, that
Majorana-Weyl spinors exist only in a Minkowski spacetime
of dimension $D=2$ (mod 8), which includes the dimension $D=10$
in which superstrings may be constructed, which is not an accident.

Given that the Dirac $\gamma^\mu$ matrices provide a representation
space of the Lorentz group, it should be possible to display
explicitly the associated generators. Indeed, it may be shown
that the latter are obtained as
\begin{equation}
\Sigma^{\mu\nu}=\frac{1}{2}i\gamma^{\mu\nu}\ \ \ ,\ \ \ 
\gamma^{\mu\nu}=\frac{1}{2}\left[\gamma^\mu,\gamma^\nu\right]\ ,
\end{equation}
with
\begin{equation}
\gamma^{\mu\nu}=\frac{1}{2}\left(\begin{array}{c c}
	\sigma^\mu\overline{\sigma}^\nu-\sigma^\nu\overline{\sigma}^\mu & 0 \\
0 & \overline{\sigma}^\mu\sigma^\nu-\overline{\sigma}^\nu\sigma^\mu
	\end{array}\right)\ .
\end{equation}
Thus a right-handed spinor $\psi_\alpha$ transforms according to the
generators,
\begin{equation}
\Sigma_R^{\mu\nu}\ :\ \ 
{\left(\Sigma_R^{\mu\nu}\right)_\alpha}^\beta=\frac{1}{4}i\left[
{\sigma^\mu}_{\alpha\dot{\gamma}}\,
\overline{\sigma}^{\nu\dot{\gamma}\beta}\,-\,
{\sigma^\nu}_{\alpha\dot{\gamma}}\,
\overline{\sigma}^{\mu\dot{\gamma}\beta}\right]\ ,
\end{equation}
while a left-handed Weyl spinor $\overline{\chi}^{\dot{\alpha}}$ according to
\begin{equation}
\overline{\Sigma}_L^{\mu\nu}\ :\ \ 
{\left(\overline{\Sigma}_L^{\mu\nu}\right)^{\dot{\alpha}}}_{\dot{\beta}}=
\frac{1}{4}i\left[
\overline{\sigma}_L^{\mu\dot{\alpha}\gamma}\,
{\sigma^\nu}_{\gamma\dot{\beta}}\,-\,
\overline{\sigma}^{\nu\dot{\alpha}\gamma}\,
{\sigma^\mu}_{\gamma\dot{\beta}}\right]\ .
\end{equation}

Given these different considerations, it should not come as a
surprise that once a free quantum field theory dynamics is constructed,
it turns out that such fundamental spinor representations of the
Lorentz group describe quanta which are massive or massless
particles whose spin or helicity is 1/2.

Extending the above considerations to an arbitrary representation
of the Dirac-Clifford algebra, any Dirac spinor may
be decomposed into its chiral components,
\begin{equation}
\psi=\psi_L+\psi_R\ \ ,\ \ 
\psi_L=P_L\,\psi=\frac{1}{2}\left(1-\gamma_5\right)\psi\ \ ,\ \ 
\psi_R=P_R\,\psi=\frac{1}{2}\left(1+\gamma_5\right)\psi\ .
\end{equation}
The SL(2,$\mathbb C$) invariant tensors that enable the raising
and lowering of dotted and undotted indices provide for
a transformation which, given a Dirac spinor $\psi$ and its
complex conjugate, constructs another Dirac spinor also transforming
according to the correct rules under Lorentz transformations.
This operation, known as charge conjugation since it exchanges the roles
played by particles and their antiparticles, is represented through
a matrix $C$ such that
\begin{equation}
C\gamma^\mu C^{-1}=-{\gamma^\mu}^{\rm T}\ \ ,\ \ 
C=i\gamma^2\gamma^0\ \ ,\ \ 
C^\dagger=C^{\rm T}=-C\ \ ,\ \ C^2=-\one\ ,
\end{equation}
where, except for the very first identity,
the last series of properties is valid, for instance,
in the Dirac and chiral representations of the $\gamma^\mu$ matrices,
but not necessarily in just any other representation of the Dirac-Clifford
algebra. The charge conjugate Dirac spinor $\psi_C$ associated to a given 
Dirac spinor $\psi$ is given by,
\begin{equation}
\psi_C=C\overline{\psi}^{\rm T}\ \ ,\ \ 
\overline{\psi}=\psi^\dagger\,\gamma^0\ ,
\end{equation}
up to an arbitrary phase factor. Consequently, a Majorana spinor $\psi$
obeys the Majorana condition,
\begin{equation}
\psi=\psi_C=C\overline{\psi}^{\rm T}\ ,
\end{equation}
thus extending to the Dirac spinor representation of the Lorentz
group in a manner consistent with Lorentz transformation, the reality
condition under complex conjugation for such fields, in a way
similar to the simple reality condition $\phi=\phi^\dagger$ for
a scalar field real under complex conjugation describing spin 0 
particles which are their own antiparticles.

Given all the above, different properties may be established.
For instance, one has
\begin{equation}
\overline{\left(\psi_L\right)}=\left(\overline{\psi}\right)_R\ ,\
\overline{\left(\psi_R\right)}=\left(\overline{\psi}\right)_L\ ,\
\left(\psi_L\right)_C=\left(\psi_C\right)_R\ ,\ 
\left(\psi_R\right)_C=\left(\psi_C\right)_L\ .
\end{equation}
Lorentz invariant spinor bilinears decompose as
\begin{equation}
\overline{\psi}\chi=\overline{\psi_L}\chi_R+\overline{\psi_R}\chi_L\ ,\
\overline{\psi}\gamma_5\chi=
\overline{\psi_L}\gamma_5\chi_R+\overline{\psi_R}\gamma_5\chi_L
=\overline{\psi_L}\chi_R-\overline{\psi_R}\chi_L\ ,
\label{eq:chiral1}
\end{equation}
where, under parity, the first quantity is a pure scalar, and
the second a pseudoscalar. Likewise, one has the Lorentz covariants,
\begin{equation}
\begin{array}{r c l}
\overline{\psi}\gamma^\mu\chi&=&
\overline{\psi_L}\gamma^\mu\chi_L+\overline{\psi_R}\gamma^\mu\chi_R\ ,\\
 & & \\
\overline{\psi}\gamma^\mu\gamma_5\chi&=&
-\overline{\psi_L}\gamma^\mu\chi_L+\overline{\psi_R}\gamma^\mu\chi_R\ ,\\
 & & \\
\overline{\psi}\sigma^{\mu\nu}\chi&=&
\overline{\psi_L}\sigma^{\mu\nu}\chi_R+
\overline{\psi_R}\sigma^{\mu\nu}\chi_L\ ,
\end{array}
\label{eq:chiral2}
\end{equation}
where in the last relation one defines 
$\sigma^{\mu\nu}=i[\gamma^\mu,\gamma^\nu]/2$.  Note that the
bilinears $\overline{\psi}\gamma^\mu\chi$,
$\overline{\psi}\gamma^\mu\gamma_5\chi$ and
$\overline{\psi}\sigma^{\mu\nu}\chi$ transform as
a 4-vector, an axial 4-vector, and a $(1,0)\oplus(0,1)$ tensor,
respectively. In fact, the whole $2^4=16$ dimensional Dirac-Clifford 
algebra, generated by the $2^2\times 2^2$ matrices $\one$ and $\gamma^\mu$, 
is spanned by the $2^4=16$ independent
quantities $\one$, $\gamma_5$, $\gamma^\mu$,
$\gamma^\mu\gamma_5$ and $\sigma^{\mu\nu}$ (one has indeed
$\sigma^{\mu\nu}\gamma_5=i\epsilon^{\mu\nu\rho\sigma}\sigma_{\rho\sigma}/2$
where $\epsilon^{0123}=+1$).

Further identities involving four Dirac spinors are also important
to establish supersymmetry invariance. These involve the
celebrated Fierz identities,\cite{QFT} the simplest of which is of the 
form,\footnote{As a matter of fact, all other Fierz identities follow 
from the present one, by appropriate choices of the spinors involved.}
\begin{equation}
\begin{array}{r l}
\overline{\psi_1}\one\psi_2\,\overline{\psi_3}\one\psi_4=
-\frac{1}{4}\Big\{&
\overline{\psi_1}\one\psi_4\,\overline{\psi_3}\one\psi_2\,+\,
\overline{\psi_1}\gamma^\mu\psi_4\,\overline{\psi_3}\gamma_\mu\psi_2\,+\,\\
 & \\
&+\frac{1}{2}\overline{\psi_1}\sigma^{\mu\nu}\psi_4\,
\overline{\psi_3}\sigma_{\mu\nu}\psi_2\,-\,\\
 & \\
& - \overline{\psi_1}\gamma^\mu\gamma_5\psi_4\,
\overline{\psi_3}\gamma_\mu\gamma_5\psi_2\,+\,
\overline{\psi_1}\gamma_5\psi_4\,\overline{\psi_3}\gamma_5\psi_2\Big\}\ ,
\end{array}
\end{equation}
where $\psi_1$, $\psi_2$, $\psi_3$ and $\psi_4$ are arbitrary
Grassmann odd Dirac spinors. An application of this identity leads,
for instance, to the relation
\begin{equation}
\overline{\epsilon_{1R}}\,\partial_\mu\psi_L\,\gamma^\mu\epsilon_{2R}=
-\frac{1}{2}\overline{\epsilon_{1R}}\gamma_\nu\epsilon_{2R}\,
\gamma^\mu\gamma^\nu\partial_\mu\psi_L\ ,
\end{equation}
where $\epsilon_{1R}$, $\epsilon_{2R}$ and $\psi_L$ are Grassmann odd
Dirac spinors of definite chirality as indicated by their lower label. 
This relation is central in establishing the supersymmetry invariance 
property of the simplest example of a supersymmetric field theory, 
the so-called Wess-Zumino model involving a scalar and a Weyl 
or Majorana spinor.\cite{WZ,Deren}

In the case of Grassmann odd Majorana spinors $\epsilon$ and $\lambda$,
one also has,
\begin{equation}
\begin{array}{r c c c l}
\overline{\epsilon}\lambda&=&\overline{\lambda}\epsilon&=&
\left(\overline{\epsilon}\lambda\right)^\dagger\ ,\\
\overline{\epsilon}\gamma_5\lambda&=&
\overline{\lambda}\gamma_5\epsilon&=&
-\left(\overline{\epsilon}\gamma_5\lambda\right)^\dagger\ ,\\
\overline{\epsilon}\gamma^\mu\lambda&=&
-\overline{\lambda}\gamma^\mu\epsilon&=&
-\left(\overline{\epsilon}\gamma^\mu\lambda\right)^\dagger\ ,\\
\overline{\epsilon}\gamma^\mu\gamma_5\lambda&=&
\overline{\lambda}\gamma^\mu\gamma_5\epsilon&=&
\left(\overline{\epsilon}\gamma^\mu\gamma_5\lambda\right)^\dagger\ ,\\
\overline{\epsilon}\gamma^\mu\gamma^\nu\lambda&=&
-\overline{\lambda}\gamma^\mu\gamma^\nu\epsilon&=&
\left(\overline{\epsilon}\gamma^\mu\gamma^\nu\lambda\right)^\dagger\ .
\end{array}
\end{equation}
It is a useful exercise to establish any of these identities.

\subsection{The Dirac Equation}
\label{Sec3.5}

Let us now consider the dynamics of a single free Dirac spinor
field, thus described, at the classical level, by complex
valued Grassmann odd variables forming a 4-component Dirac
spinor $\psi(x^\mu)$. The action principle for
a such a system is given by the Lorentz invariant quantity
\begin{equation}
S\left[\psi,\overline{\psi}\right]=\int d^4x^\mu\,
{\mathcal L}\left(\psi,\partial_\mu\psi\right)\ ,
\end{equation}
with the Lagrangian density\footnote{Often, this Lagrangian density
is given as ${\mathcal L}=i\overline{\psi}\gamma^\mu\partial_\mu\psi-
m\overline{\psi}\psi$, which differs from the one given here by a total
divergence with no consequence for a choice of boundary conditions
at infinity such that fields vanish asymptotically. Note however
that the form chosen in (\ref{eq:DiracL}) is manifestly real under
complex conjugation, as befits any Lagrangian density.}
\begin{equation}
{\mathcal L}=\frac{1}{2}i\left[\,\overline{\psi}\gamma^\mu\partial_\mu\psi\,-\,
\partial_\mu\overline{\psi}\gamma^\mu\psi\,\right]\,-\,m\overline{\psi}\psi\ .
\label{eq:DiracL}
\end{equation}
Through the variational principle, the associated equation of motion
is the celebrated Dirac equation,
\begin{equation}
\left[i\gamma^\mu\partial_\mu\,-\,m\right]\,\psi(x^\mu)=0\ .
\label{eq:Dirac}
\end{equation}

A few remarks are in order. Given the relations in (\ref{eq:chiral1})
and (\ref{eq:chiral2}), it is clear that the kinetic term
$\overline{\psi}\gamma^\mu\partial_\mu\psi$ couples the chiral
components of the Dirac spinor by preserving their chirality,
while the coupling $m\overline{\psi}\psi$ switches between the
two chirality components. As will become clear hereafter, since
the real parameter $m\ge 0$ in fact determines the mass of the particle quanta
associated to such a field, a massless Dirac particle propagates
without flipping its chirality, whereas a massive particle sees both
its chiral components contribute to its spacetime dynamics.

The term $m\overline{\psi}\psi$ is known as the Dirac mass term.
In particular, it preserves the symmetry of the kinetic term under
global phase transformations of the Dirac spinor, 
\begin{equation}
{\rm U}_{\rm V}{\rm (1)}\ :\ \ \ 
\psi'(x)=e^{i\alpha}\,\psi(x)\ ,
\end{equation}
leading to a conserved U$_{\rm V}$(1) quantum number which, effectively, 
counts the difference between the numbers of fermions and antifermions
present in the system. This U$_{\rm V}$(1) phase symmetry is thus that of
the fermion number, which may coincide with the electric charge
quantum number when coupled to the electromagnetic interaction.
The corresponding conserved Noether current is simply the vector
bilinear $J^\mu=\overline{\psi}\gamma^\mu\psi$, thus obeying the
divergenceless condition $\partial_\mu J^\mu=0$ for solutions to
the Dirac equation (\ref{eq:Dirac}).

Furthermore, since under the transformation
\begin{equation}
\psi'(x)=\gamma_5\,\psi(x)\ ,
\end{equation}
the mass term $m\overline{\psi}\psi$ changes sign,
$\overline{\psi'}\psi'=-\overline{\psi}\psi$, it may always be
assumed that the parameter $m$ is not negative, $m\ge 0$.

One may also consider U(1)$_{\rm A}$ axial transformations,
\begin{equation}
{\rm U}_{\rm A}{\rm (1)}\ :\ \ \ 
\psi'(x)=e^{i\alpha\gamma_5}\,\psi(x)\ ,
\end{equation}
leaving the kinetic term of the Lagrangian density invariant, but
not the Dirac mass term. When $m=0$, the associated conserved
Noether current density is the axial vector spinor bilinear,
$J^\mu_5=\overline{\psi}\gamma^\mu\gamma_5\psi$, which is indeed
conserved for solutions to Dirac's equation (\ref{eq:Dirac})
only provided $m=0$, as may explicitly be checked through
direct calculation. These vector and axial symmetries of the
Dirac Lagrangian density are important aspects for the theory
of the strong interactions, quantum chromodynamics (QCD).

Rather than considering a Dirac mass term, one may also use
the charge conjugate spinor $\psi_C$ to define another type of mass term,
\begin{equation}
m_M\overline{\psi}\psi_C\,+\,{\rm hermitian\ conjugate}\ ,
\end{equation}
known as a Majorana mass term. However, it should be clear that
such a term breaks not only the axial symmetry as does a Dirac
mass term, but also the above vector symmetry under phase transformations.
Hence, a Majorana mass term leads to a violation of the fermion
number, again a reason why such a possibility may be contemplated
for neutrinos only within the Standard Model of the quarks and leptons
and their strong and electroweak interactions.

A detailed analysis, similar to that applied to the Klein--Gordon
equation,\cite{GovCOPRO2} considering the plane wave solutions\footnote{Such
solutions must exist since the Dirac equation is invariant under
spacetime translations and is linear in the field.} to the Dirac 
equation (\ref{eq:Dirac}), reveals that the general solution may 
be expressed through the following mode expansion
\begin{equation}
\psi(x^\mu)=\int\frac{d^3\vec{k}}{(2\pi)^32\omega(\vec{k}\,)}\,
\sum_{s=\pm}\left\{
e^{-ik\cdot x}\,u(\vec{k},s)b(\vec{k},s)\,+\,
e^{ik\cdot x}\,v(\vec{k},s)d^\dagger(\vec{k},s)\right\}\ ,
\label{eq:solution}
\end{equation}
where the plane wave spinors $u(\vec{k},s)$ and $v(\vec{k},s)$
are positive- and negative-frequency
solutions to the Dirac equation in energy-momentum space,
\begin{equation}
\left[\gamma^\mu k_\mu-m\right]\,u(\vec{k},s)=0\ \ \ ,\ \ \ 
\left[\gamma^\mu k_\mu+m\right]\,v(\vec{k},s)=0\ .
\end{equation}
The normalisation of these spinors is such that
\begin{equation}
\sum_{s=\pm}\,u(\vec{k},s)\overline{u}(\vec{k},s)=
\left(\gamma^\mu k_\mu+m\right)\ \ \ ,\ \ \ 
\sum_{s=\pm}\,v(\vec{k},s)\overline{v}(\vec{k},s)=
\left(\gamma^\mu k_\mu-m\right)\ .
\end{equation}
The index $s=\pm$ taking two values is related to a spin or a helicity
projection degree of freedom, specifying the polarisation state of the
solution. The general solution has to include a summation over the
two possible polarisation states of the field. The spinors $u(\vec{k},s)$
and $v(\vec{k},s)$ thus also correspond to polarisation spinors
characterising the polarisation state of the field (in the same way that
a polarisation vector characterises the polarisation state of a vector
field $A_\mu(x^\mu)$, such as the electromagnetic vector field).

Finally, in exactly the same manner as for the scalar field,\cite{GovCOPRO2}
the quantities $b(\vec{k},s)$ and $d^\dagger(\vec{k},s)$ are, at
the classical level, Grassmann odd integration constants specifying 
a unique solution to the Dirac equation, which, at the quantum level, 
correspond to quantum operators for which the quantum algebraic structure 
is given by the following anticommutation relations
\begin{equation}
\left\{b(\vec{k},s),b^\dagger(\vec{\ell},r)\right\}=
(2\pi)^32\omega(\vec{k}\,)\,\delta_{sr}\,\delta^{(3)}(\vec{k}-\vec{\ell}\,)=
\left\{d(\vec{k},s),d^\dagger(\vec{\ell},r)\right\}\ .
\end{equation}
Note that the normalisation of these relations is the same as that of
the creation and annihilation operators for a scalar field. As
explained in Ref.~\refcite{GovCOPRO2}, this choice leads a Lorentz
covariant normalisation of 1-particle states and mode decomposition
of fields.

One very important point should be emphasized here. By giving the
above anticommutation relations, it is understood, as it is also in
the bosonic case, that only the nonvanishing anticommutators are
displayed. Thus the following anticommutation relations are
implicit,
\begin{equation}
\begin{array}{r c l}
\left\{b(\vec{k},s),b(\vec{\ell},r)\right\}=&0&=
\left\{d(\vec{k},s),d(\vec{\ell},r)\right\}\ ,\\
 & & \\
\left\{b^\dagger(\vec{k},s),b^\dagger(\vec{\ell},r)\right\}=&0&=
\left\{d^\dagger(\vec{k},s),d^\dagger(\vec{\ell},r)\right\}\ .
\end{array}
\label{eq:vanish}
\end{equation}
Given that $b(\vec{k},s)$ and $d(\vec{k},s)$ are to be
interpreted as annihilation ope\-ra\-tors for particles and antiparticles,
and $b^\dagger(\vec{k},s)$ and $d^\dagger(\vec{k},s)$ as creation
operators for particles and antiparticles, respectively, the
anticommutators in (\ref{eq:vanish}) have as consequence that
no two identical particles may occupy the same quantum state
specified by the quantum numbers $\vec{k}$ and $s$. In other words,
in contradistinction to commutation relations for bosonic degrees
of freedom as is the case for a scalar field, anticommutation relations
provide a manifest realisation of the Pauli exclusion principle at
the operator level. Subsequent action with the same creation operator
on a 1-particle state, $|\vec{k},s;-\rangle=b^\dagger(\vec{k},s)|0\rangle$
or $|\vec{k},s;+\rangle=d^\dagger(\vec{k},s)|0\rangle$, leads to
the null vector in Hilbert space, 
since ${b^\dagger}^2(\vec{k},s)=0={d^\dagger}^2(\vec{k},s)$. 
It thus appears that half-integer spin
fields, namely fermionic degrees of freedom, must be quantised
according to anticommutation relations, whereas integer spin fields,
namely bosonic degrees of freedom, must be quantised with commutation
relations. This is the realisation of the spin-statistics connection.

The justification of this choice may be seen from a series of arguments.
The one often invoked goes as follows.\cite{QFT} Given the different mode 
expansions of the bosonic and fermionic fields in terms of creation and 
annihilation operators in a Fock space representation of their Fock algebra, 
it is necessary to specify an ordering prescription for composite operators,
such as for instance the Hamiltonian operator measuring the total energy 
content of the field. Within the perturbative Fock space representation, it is
customary and natural to choose normal ordering, whereby all
creation operators are brought to sit to the left of all annihilation
operators. In the case of the Dirac spinor though, when using
commutation relations rather than anticommutation ones, this
prescription leads to an energy spectrum which is not bounded below:
the contribution of the $d^\dagger d$ type (antiparticles) is
negative-definite! On the other hand, using anticommutation relations
brings in the required minus sign, rendering the energy spectrum of
the system positive-definite both for particles and antiparticles.
Half-integer spin fields must be quantised according to anticommutation
relations.

For that reason, it is also necessary to use at the classical level
Grassmann odd degrees of freedom to describe half-integer spin systems.
Consequently, the usual Hamiltonian formulation of such systems
involves now Grassmann graded Poisson brackets,\cite{GovBook} extending 
the properties of the usual bosonic Poisson brackets based on commuting 
degrees of freedom, as is the case for the scalar field for instance. 
Through the correspondence principle, such Grassmann graded Poisson 
brackets must then correspond to Grassmann graded (anti)commutation 
relations for the quantised system, in particular anticommutation 
relations for fermionic degrees of freedom of half-integer spin and 
commutation relations for bosonic degrees of freedom of integer spin. 
The algebraic properties shared by Grassmann graded Poisson brackets and
Grassmann graded (anti)commutation relations are indeed identical,
hence the necessity of such a coherent prescription for their correspondence.

{}From yet another point of view, the necessity of Grassmann odd degrees
of freedom for spinor fields may be seen as follows. Note that the
Lagrangian function for the Dirac field is linear in the spacetime
gradient $\partial_\mu\psi$, whereas that for the scalar field is
quadratic in $\partial_\mu\phi$.\footnote{This also means that the
Dirac Lagrangian is already in Hamiltonian form.\cite{GovBook,FJ}} 
This is a crucial fact, when considered in relation to the possibility 
of adding total derivatives to Lagrange functions. Indeed, for the sake 
of the argument, consider a one degree of freedom system of configuration 
space coordinate $\theta(t)$, for which the Lagrange function 
is first-order in the time derivative,
\begin{equation}
L=N\theta\frac{d\theta}{dt}-V(\theta)\ ,
\end{equation}
$N$ being some normalisation constant with properties under complex
conjugation such that $L$ be real ($\theta$ could be complex valued).
However, one may also write
\begin{equation}
\theta\frac{d\theta}{dt}=\frac{d}{dt}\left(\theta^2\right)\,-\,
\frac{d\theta}{dt}\theta\ .
\end{equation}
Thus, if the variable $\theta$ is Grassmann even, namely implying that
$\theta$ and $\dot{\theta}$ commute, one has
\begin{equation}
\theta\frac{d\theta}{dt}=\frac{d}{dt}\left(\frac{1}{2}\theta^2\right)\ ,
\end{equation}
showing that such a first-order contribution to such an
action for a bosonic degree of freedom reduces purely to a total time 
derivative, hence leads to an equation of motion
which is not a dynamical equation but rather a constraint condition,
$\partial_\theta V(\theta)=0$,
involving only the $\theta$-derivative of the potential contribution
$V(\theta)$ to the Lagrange function.
On the other hand, if the variable $\theta$ is Grassmann odd, namely
such that $\theta^2=0$ and $\dot{\theta}\theta=-\theta\dot{\theta}$
(since $\theta_1\theta_2=-\theta_2\theta_1$ for
Grassmann odd variables $\theta_1$ and $\theta_2$),
the first-order contribution $\theta\dot{\theta}$ to the Lagrange 
function does indeed lead to an equation of motion describing dynamics, 
namely
\begin{equation}
\dot{\theta}=\frac{1}{2N}\frac{\partial V}{\partial\theta}\ ,
\end{equation}
where in the r.h.s. a left-derivative is implicitly understood.
Hence, first-order actions of the above type, which generically apply
for spinor field representations of the Lorentz group,
need to be defined in terms of Grassmann odd variables in order to
lead to nontrivial dynamics. Consequently, at the quantum level,
they need to be quantised using anticommutation, rather than commutation
relations. 

The whole mathematical framework is thus consistent,
both at the classical as well as the quantum level, provided
integer spin degrees of freedom are described in terms of bosonic
or commuting Grassmann even variables, hence commutation relations at 
the quantum level, and half-integer spin degrees of freedom are described 
in terms of fermionic or anticommuting Grassmann odd variables, hence
anticommutation relations at the quantum level.

Having understood how to quantise the Dirac spinor field, let us conclude
with a few more remarks. First, consider the Majorana condition
$\psi_C(x)=\psi(x)$ imposed on such a spinor. The associated Lagrangian
density then reads,
\begin{equation}
\begin{array}{r c l}
{\mathcal L}&=&\frac{1}{4}i\left[\overline{\psi}\gamma^\mu\partial_\mu\psi\,-\,
\partial_\mu\overline{\psi}\gamma^\mu\psi\right]-
\frac{1}{2}m\overline{\psi}\psi\\
 & & \\
&=&\frac{1}{2}i\overline{\psi}\gamma^\mu\partial_\mu\psi\,-\,
\frac{1}{2}m\overline{\psi}\psi+
\partial_\mu\left(-\frac{1}{4}i\overline{\psi}\gamma^\mu\psi\right)\ ,
\end{array}
\end{equation}
where the choice of factor $1/2$ in comparison to the Dirac Lagrangian
density is made in order to have a convenient normalisation of
the field, leading to the usual normalisation of the anticommutation
relations for the creation and annihilation operators of its quanta.
This factor is also related to the avoidance of double counting
of degrees of freedom. In fact, it is the same factor\cite{GovCOPRO2} that
appears in the Lagrangian density for a real scalar field, as
compared to that for a complex scalar field $\phi(x)$, namely
related to the factor $1/\sqrt{2}$ in the real and imaginary
components of the complex field in terms of real fields,
$\phi(x)=(\phi_1(x)+i\phi_2(x))/\sqrt{2}$.

Solving the Dirac equation following from the above Majorana field 
Lagrangian density, subject to the Majorana condition, leads to the
mode decomposition,
\begin{equation}
\psi(x)=\int\frac{d^3\vec{k}}{(2\pi)^32\omega(\vec{k}\,)}
\sum_{s=\pm}\left\{
e^{-ik\cdot x}\,u(\vec{k},s)b(\vec{k},s)\ +\
e^{ik\cdot x}\,v(\vec{k},s)b^\dagger(\vec{k},s)\right\}\ ,
\end{equation}
with the same quantities as those that appear in the solution
(\ref{eq:solution}) for the Dirac spinor. Note well that indeed
there no longer appears the creation operator $d^\dagger(\vec{k},s)$ for
antiparticles, but that only the annihilation, $b(\vec{k},s)$,
and creation, $b^\dagger(\vec{k},s)$, operators of particles of
a single type contribute to the Majorana spinor field 
operator.\footnote{Again, this conclusion is in perfect analogy with
what happens for a real and a complex scalar field.\cite{GovCOPRO2}}
A Majorana spinor describes quanta which are their own antiparticles.
Hence, they cannot carry a conserved quantum number, such as
fermion number, as was already observed previously. A Majorana spinor
describes neutral spin 1/2 particles, whereas a Dirac spinor
describes charged (for some symmetry, for instance the U(1) symmetry
of electric charge or fermion number) spin 1/2 particles.

The fermion number of the Dirac spinor is determined, through
Noether's theorem, from the time component of the conserved
vector current $J^\mu=\overline{\psi}\gamma^\mu\psi$. In terms of
the mode expansion, one has
\begin{equation}
F=\int_{(\infty)}d^3\vec{x}\,J^0=
\int\frac{d^3\vec{k}}{(2\pi)^3\omega(\vec{k}\,)}\sum_{s=\pm}\left\{
b^\dagger(\vec{k},s)b(\vec{k},s)\,-\,d^\dagger(\vec{k},s)d(\vec{k},s)\right\}\ ,
\end{equation}
where the normal ordering prescription has been applied. Clearly, this
expression shows that states created by $b^\dagger(\vec{k},s)$ carry
an $F$ value opposite to that carried by states created by
$d^\dagger(\vec{k},s)$. The conserved $F$ quantum number, related to
the invariance of the Dirac Lagrangian density under arbitrary global
phase transformations of the Dirac spinor field, is what distinguishes
particles from antiparticles of spin 1/2 in this system. If this
quantum number is also identified to the electric charge of the
electromagnetic interaction for electrons, it is thus seen that
the Dirac spinor describes both electrons and their antiparticles,
positrons, of identical mass and spin, but opposite electric charge,
which remains a conserved quantum number. Gauging the associated
U$_{\rm V}$(1) vector symmetry then leads to a complete description of the
quantum electromagnetic interactions between electrons, positrons and photons,
namely quantum electrodynamics (QED). When this is extended to
nonabelian internal symmetries, one obtains Yang--Mills 
theories\cite{GovCOPRO2} which, for the choice of gauge group 
SU(3)$_{\rm C}\times$SU(2)$_{\rm L}\times$U(1)$_{\rm Y}$,
enter the construction of the Standard Model of quarks and leptons
and their interactions. The sector of the strong interactions among quarks
is thus based on the colour symmetry SU(3)$_{\rm C}$ and the associated
Yang--Mills gauge theory of quantum chromodynamics (QCD).

For what concerns spacetime symmetries, the Poincar\'e generators
are now given by the expressions,\cite{QFT}
\begin{equation}
P^\mu=\int\frac{d^3\vec{k}}{(2\pi)^3\omega(\vec{k}\,)}\,k^\mu\,
\sum_{s=\pm}\left\{
b^\dagger(\vec{k},s)b(\vec{k},s)\,+\,d^\dagger(\vec{k},s)d(\vec{k},s)\right\}\ ,
\end{equation}
\begin{equation}
M^{\mu\nu}=\int_{(\infty)} d^3\vec{x}\left\{
\Theta^{0\mu}x^\nu-\Theta^{0\nu}x^\mu-
\frac{1}{4}\overline{\psi}\left\{\gamma^0,\sigma^{\mu\nu}\right\}\psi\right\}\ ,
\end{equation}
where
\begin{equation}
\Theta^{\mu\nu}=\frac{1}{2}i\left[\overline{\psi}\gamma^\mu\partial^\nu\psi\,-\,
\partial^\nu\overline{\psi}\gamma^\mu\psi\right]\ ,
\end{equation}
while fermion normal ordering is implicit of course.
It then follows that the 1-particle states obtained by acting with
the creation operators $b^\dagger(\vec{k},s)$ and $d^\dagger(\vec{k},s)$
on the Fock vacuum $|0\rangle$ are energy-momentum eigenstates of
momentum $\vec{k}$ and mass $m$, possessing spin or helicity 1/2.

In the same way as for the scalar field,\cite{GovCOPRO2} it is
possible to compute the Feynman propagator of the Dirac field,
namely the causal probability amplitude for seeing a particle created
at a given point in spacetime and annihilated at some other such
point. This time-ordered amplitude is thus defined by the 2-point
correlation function
\begin{equation}
\begin{array}{r c l}
\langle 0|T\psi_\alpha(x)\overline{\psi}_\beta(y)|0\rangle&=&
\theta(x^0-y^0)\langle 0|\psi_\alpha(x)\overline{\psi}_\beta(y)|0\rangle\,-\,\\
 & & \\
&&-\theta(y^0-x^0)\langle 0|\overline{\psi}_\beta(y)\psi_\alpha(x)|0\rangle\ ,
\end{array}
\end{equation}
where the anticommuting nature of the spinor is accounted for
through the negative sign in the second contribution in the r.h.s.
($\theta(x)$ denotes the usual step function, $\theta(x>0)=1$ and
$\theta(x<0)=0$). In the case of the free Dirac field, a direct 
substitution of the mode expansion (\ref{eq:solution}) leads to 
the integral representation,
\begin{equation}
\langle 0|T\psi_\alpha(x)\overline{\psi}_\beta(y)|0\rangle=
\int\frac{d^4k}{(2\pi)^4}\,e^{-ik\cdot(x-y)}\,
\left(\frac{i}{\gamma^\mu k_\mu-m+i\epsilon}\right)_{\alpha\beta}\ ,
\end{equation}
where, as usual,\cite{GovCOPRO2} $\epsilon>0$ corresponds to 
an infinitesimal imaginary part
in the denominator of the momentum-space propagator introduced 
to specify the contour integration in the complex $k^0$ energy plane 
in order to pick up the correct pole contributions associated to 
the positive- and negative-frequency components of the Dirac spinor 
mode expansion. This Dirac propagator is the basis for perturbation theory
involving Dirac spinors, in the same way that the Feynman propagator
for scalar fields enables the evaluation of the perturbation theory corrections
stemming from interactions between scalar particles.\cite{GovCOPRO2}

\section{On the Road Towards Supersymmetry:\\
A Simple Quantum Mechanical Model}
\label{Sec4}

The previous sections have reviewed how, by enforcing at all
steps the consequences of spacetime Lorentz and Poincar\'e covariance,
relativistic quantum field theories lead to a conceptual framework
deeply rooted in basic physical principles which naturally describes
the relativistic and quantum properties of point-particles of
given mass and spin, and the possibility of their creation and
annihilation in a variety of processes for which the fundamental
interactions are responsible. The Poincar\'e symmetry invariance properties
of Minkowski spacetime allow for the particle interpretation of
definite energy-momentum and spin values for the quanta of such fields.
Any further internal symmetries then also account for further
conserved quantum numbers that particles carry. When gauged,
such internal symmetries lead to specific interactions of the
Yang--Mills type, which are at the basis of the construction
of the successful Standard Model for quarks and leptons and
their fundamental interactions.

We have also made clear how bosonic particles of integer spin need
to be described in terms of commuting degrees of freedom and
quantum commutation relations for the tensor field representations of the
Lorentz group, whereas fermionic particles of half-integer spin need
to be described in terms of anticommuting degrees of freedom
and quantum anticommutation relations for the spinor field
representations of the Lorentz group.

As briefly discussed in Ref.~\refcite{GovCOPRO2}, this widely 
encompassing framework aiming towards a fundamental unification 
has now come to a cross-roads at which an irreconcilable clash 
has arisen between the principles of general relativity, the 
relativistic invariant classical field theory for the gravitational 
interaction described through the dynamics of spacetime geometry, 
and the principles of relativistic quantum field theory, the natural framework 
for all of matter and the other three fundamental interactions.
Many extensions beyond the Standard Model aiming at a resolution
of this conflict have been contemplated, most of which involve
in one way or another algebraic structures relating fermionic and
bosonic degrees of freedom, so-called supersymmetry algebras.
Indeed, the distinct separation between boson and fermionic fields
at the same time attracts the suggestion of a possible unification
within a larger framework in which such degrees of freedom could
appear on an equal footing, a specific type of a fundamental unification
of matter (half-integer spin particles, namely the quarks and leptons) 
and interactions (integer spin particles, namely the Yang--Mills gauge bosons
of the strong and electroweak interactions, the higgs particle yet to be
discovered and the graviton). One should expect that assuming this 
to be achievable, such a unification should also extend the usual 
commuting coordinates of Minkowski spacetime into a superspace 
including both commuting and anticommuting coordinates, truly a first 
embodiement of an eventual fundamental quantum geometry.

The stage has thus been set to embark onto a journey on the roads
towards the construction of supersymmetric quantum field
theories. These notes shall stop short of such a discussion,
which is widely available in the literature, and conclude in this section
with a series of remarks pointing towards the generic features of
such systems, as a way of opening the reader's mind for whom this is
unknown territory of theoretical physics, to what he/she may expect
from a study on his/her own of supersymmetry.

We shall do this starting again from ordinary quantum mechanics.
Hopefully, it should have been made abundantly clear\cite{GovCOPRO2}
that the ``essence'' of relativistic quantum fields is their
harmonic oscillator characteristics, extended in such a manner as to
make their spacetime dynamics also consistent with the Poincar\'e
invariance of Minkowski spacetime. This is true whether for bosonic
or fermionic quantum fields, the simplest examples of which are
the fields describing particles of spin or helicity 0 and 1/2.
Let us thus reduce to the extreme again these field situations,
by restricting the discussion to simple harmonic oscillator degrees
of freedom finite in number. The generalisation to field degrees of
freedom will then be restricted and guided by the constraints stemming
from Poincar\'e invariance, leading {\it in fine\/} to supersymmetric
relativistic quantum field theories.

To begin with, let us consider a single bosonic harmonic 
oscillator.\cite{GovCOPRO2} Once quantised, to such a system is
associated a representation space of its quantum states, its physical
Hilbert space, on which act the annihilation, $a$, and creation, $a^\dagger$,
operators of energy quanta subjected to the commutation relation
$[a,a^\dagger]=\one$ (the other commutators vanish identically,
$[a,a]=0=[a^\dagger,a^\dagger]$). A canonical basis of the Fock algebra is
the Fock basis, constructed from a vacuum state $|0\rangle$
annihilated by $a$, $a|0\rangle=0$, on which acts the creation
operator $a^\dagger$, leading to the discrete set of states 
$|n\rangle=(a^\dagger)^n|0\rangle/\sqrt{n!}$ ($n=0,1,2,\cdots$)
obeying the properties,
\begin{equation}
\langle n|m\rangle=\delta_{nm}\ \ ,\ \ 
a|n\rangle=\sqrt{n}|n-1\rangle\ \ ,\ \ 
a^\dagger|n\rangle=\sqrt{n+1}|n+1\rangle\ \ ,\ \ 
a^\dagger a|n\rangle=n|n\rangle\ .
\end{equation}

The quantum Hamiltonian of the system, generating also its dynamical
evolution in time, is diagonal in the Fock basis, and is given by
\begin{equation}
H_B=\frac{1}{2}\hbar\omega\left\{a^\dagger,a\right\}=
\frac{1}{2}\hbar\omega\left[a^\dagger a+a a^\dagger\right]=
\hbar\omega\left[a^\dagger a+\frac{1}{2}\right]\ ,
\end{equation}
where the vacuum quantum energy contribution $\hbar\omega/2$ has been
retained, while $\omega$ denotes the angular frequency of the system,
setting its energy scale in combination with Planck's constant $\hbar$.
The energy spectrum is thus equally spaced in steps of $\hbar\omega$, 
with the eigenvalues $E_B(n)=\hbar\omega(n+1/2)$, 
$H_B|n\rangle=E_B(n)|n\rangle$, starting with the vacuum state at
$E_B(n=0)=\hbar\omega/2$.

Let us now consider likewise the quantum fermionic oscillator of same angular
frequency $\omega$ (the reason being that later on we shall introduce 
a symmetry relating the bosonic and fermionic systems). The space of states
provides a representation for the fermionic anticommutator algebra
\begin{equation}
\{b,b\}=0=\{b^\dagger,b^\dagger\}\ \ \ ,\ \ \ 
\{b,b^\dagger\}=\one\ ,
\end{equation}
where $b$ and $b^\dagger$ are the fermionic annihiliation and creation
operators, respectively. Note that by having replaced commutation relations
with anticommutation ones, the vanishing anticommutators in fact imply
the properties $b^2=0={b^\dagger}^2$, the manifest realisation of the Pauli
exclusion principle for fermions. As a consequence, the Fock space
representation of this fermionic Fock algebra is 2-dimensional
(to be contrasted with the discrete infinite dimension of the bosonic
Hilbert space), and is spanned by a vacuum state $|0\rangle$ and its 
first excitation $|1\rangle=b^\dagger|0\rangle$, with the properties,
\begin{equation}
\begin{array}{r c l}
b|0\rangle=0\ \ ,\ \ 
b^\dagger|0\rangle=|1\rangle\ \ &,&\ \ 
b|1\rangle=|0\rangle\ \ ,\ \ 
b^\dagger|1\rangle=0\ ,\\
 & \\
\langle 0|0\rangle=1=\langle 1|1\rangle\ \ &,&\ \ 
\langle 0|1\rangle=0=\langle 1|0\rangle\ .
\end{array}
\end{equation}
For the quantum Hamiltonian, we shall also choose
\begin{equation}
H_F=\frac{1}{2}\hbar\omega\left[b^\dagger,b\right]=
\frac{1}{2}\hbar\omega\left[b^\dagger b-b b^\dagger\right]=
\hbar\omega\left[b^\dagger b - \frac{1}{2}\right]\ ,
\end{equation}
where this time the vacuum quantum energy is negative because of
the fermionic character of the degree of freedom. The Fock state basis
diagonalises this operator, with the energy spectrum,
\begin{equation}
H_F|0\rangle=-\frac{1}{2}\hbar\omega\ \ \ ,\ \ \ 
H_F|1\rangle=\frac{1}{2}\hbar\omega\ ,
\end{equation}
thus describing a 2-level quantum system split by an energy $\hbar\omega$.

Let us now combine these two systems, and consider the tensor product
of their operator algebras and representation spaces. Hence, the complete
Hilbert space is spanned by the states $|n,0\rangle$ and $|n,1\rangle$,
where the first entry stands for the bosonic excitation level, and the
second entry for that of the fermionic sector. The total Hamiltonian of
the system then reads,
\begin{equation}
H=H_B+H_F=\frac{1}{2}\hbar\omega\left[a^\dagger a+a a^\dagger +
b^\dagger b - b b^\dagger\right]=
\hbar\omega\left[a^\dagger a+ b^\dagger b\right]\ ,
\end{equation}
in which the vacuum quantum energies of the bosonic and fermionic sectors
have cancelled one another. Consequently, the energy eigenspectrum is
still equally spaced in steps of $\hbar\omega$, is doubly degenerate at
each level with the states $|n,0\rangle$ and $|n-1,1\rangle$ at level $n$
of energy $\hbar\omega n$, except for the single ground state or 
vacuum state $|n=0,0\rangle$ at level $n=0$ whose energy vanishes identically,
\begin{equation}
H|n=0,0\rangle=0\ \ ,\ \ 
H|n,0\rangle=\hbar\omega n|n,0\rangle\ \ ,\ \ 
H|n-1,1\rangle=\hbar\omega n|n-1,1\rangle\ .
\end{equation}

With these simple remarks, in fact we already encounter a series of
features quite unique to supersymmetry. If a system possesses a symmetry
that relates fermionic and bosonic degrees of freedom, there are 
general classes of cancellations between quantum fluctuations and corrections
stemming from the two sectors, leading to better behaved short-distance 
UV divergences generic of 4-dimensional quantum field theories. Indeed, 
there are even certain classes of quantum operators which, in supersymmetric 
field theories, are not at all renormalised by perturbative quantum corrections,
leading to very powerful so-called no-renormalisation theorems.
In addition, the cancellation between bosonic and fermionic vacuum
quantum energy contributions implies that in field theories in which
supersymmetry is not spontaneously broken, the vacuum state possesses
an exactly va\-ni\-shing energy, suggesting a possible connection with the
famous problem of the extremely small (in comparison to the Planck energy
scale relevant to quantum gravity, $10^{19}$~GeV) and yet not exactly
vanishing cosmological constant of our universe.\cite{Lambda} If only 
from that perspective, dynamical spontaneous symmetry breaking of 
supersymmetry is thus an extremely fascinating issue in the quest for a 
fundamental unification.\cite{Witten2}

The degeneracy between the bosonic states $|n,0\rangle$ and the
fermionic ones $|n,1\rangle$ suggests that there exists a
symmetry --- a supersymmetry --- re\-la\-ting these two sectors of the
system. We need to construct the operators generating such transformations,
by creating a fermion and annihilating a boson, or vice-versa, thus
mapping between bosonic and fermionic states degenerate in energy. Clearly
these operators are given by
\begin{equation}
Q=\sqrt{\hbar\omega}\ a^\dagger b\ \ \ ,\ \ \ 
Q^\dagger=\sqrt{\hbar\omega}\ ab^\dagger\ ,
\end{equation}
acting as
\begin{equation}
\begin{array}{r c l}
Q|n,0\rangle=0\ \ &,&\ \ 
Q|n,1\rangle=\sqrt{\hbar\omega}\ \sqrt{n+1}|n+1,0\rangle\ ,\\
 & & \\
Q^\dagger|n,0\rangle=\sqrt{\hbar\omega}\ \sqrt{n}|n-1,1\rangle\ \ &,&\ \ 
Q^\dagger|n,1\rangle=0\ .
\end{array}
\end{equation}
Note that the vacuum $|n=0,0\rangle$ is the single state which is
annihilated by both $Q$ and $Q^\dagger$, as it must since it is not
degenerate in energy with any other state. The operators $Q$ and $Q^\dagger$
are thus the generators of a supersymmetry present in this system.
Their algebra is given by
\begin{equation}
\left\{Q,Q\right\}=0=\left\{Q^\dagger,Q^\dagger\right\}\ \ ,\ \ 
\left\{Q,Q^\dagger\right\}=H\ \ ,\ \ 
\left[Q,H\right]=0=\left[Q^\dagger,H\right]\ .
\label{eq:SUSYalgebra}
\end{equation}
The fact that they define a symmetry is confirmed by their
vanishing commutation relations with the Hamiltonian $H$.

Once again, we uncover here a general feature of supersymmetry
algebras, namely the fact that acting twice with a supersymmetry generator,
in fact one gets an identically vanishing result, $Q^2=0={Q^\dagger}^2$,
a property directly reminiscent of cohomology classes of differential
forms in differential geometry.\cite{Witten3} In addition, the anticommutator
of a supersymmetry generator with its adjoint gives the Hamiltonian
of the system. In a certain sense thus, making a system supersymmetric
amounts to taking a square-root of its Hamiltonian. Put differently,
the square-root of the Klein--Gordon equation is the Dirac equation,
when this correspondence is extended to field theories.
{}From these simply remarks it already transpires that supersymmetry 
algebras provide powerful new tools with which to explore mathematics 
questions within a context which may draw on a lot of insight 
and intuition from quantum physics.\cite{MATH,Witten3} Results have 
indeed been very rewarding already, and many more are still to be 
established along such lines.

To complete the algebraic relations in (\ref{eq:SUSYalgebra}), it is
also useful to display the supersymmetry action on the creation and
annihilation operators,
\begin{equation}
\begin{array}{r c l}
[Q,a]=-\sqrt{\hbar\omega}\,b\ \ ,\ \ 
[Q,a^\dagger]=0\ \ &,&\ \ 
[Q^\dagger,a]=0\ \ ,\ \ 
[Q^\dagger,a^\dagger]=\sqrt{\hbar\omega}\,b^\dagger\ ,\\
 & \\
\left\{Q,b\right\}=0\ \ ,\ \ 
\left\{Q,b^\dagger\right\}=\sqrt{\hbar\omega}\,a^\dagger\ \ &,&\ \
\left\{Q^\dagger,b\right\}=\sqrt{\hbar\omega}\,a\ \ ,\ \
\left\{Q^\dagger,b^\dagger\right\}=0\ .
\end{array}
\label{eq:QFock}
\end{equation}

The properties $Q^2=0={Q^\dagger}^2$ also suggest that it should be
possible to obtain wave function representations of the fermionic
and supersymmetry algebras using complex valued Grassmann odd variables
$\theta$, such that $\theta_1\theta_2=-\theta_2\theta_1$ and thus
$\theta^2_1=0=\theta^2_2$, in the same
way that the bosonic Fock algebra possesses wave function representations
in terms of commuting coordinates, a configuration space coordinate $x$
and its conjugate momentum $p$, obeying the Heisenberg algebra.\cite{GovCOPRO2}
In the latter case, these two variables may be combined into a single
complex commuting variable $z$, leading for instance to the 
usual holomorphic representation in the bosonic sector,
\begin{equation}
a=\frac{\partial}{\partial z}\ \ \ ,\ \ \ a^\dagger=z\ .
\end{equation}
Thus likewise for the fermionic algebra, let us take
\begin{equation}
b=\frac{\partial}{\partial\theta}\ \ \ ,\ \ \ b^\dagger=\theta\ ,
\end{equation}
where it is understood that all derivatives with respect to Grassmann
odd variables are taken from the left (left-derivatives). Consequently
the supersymmetry generators are represented by
\begin{equation}
Q=\sqrt{\hbar\omega}\,z\frac{\partial}{\partial\theta}\ \ \ ,\ \ \ 
Q^\dagger=\sqrt{\hbar\omega}\,\theta\frac{\partial}{\partial z}\ ,
\end{equation}
leading to the representation for the Hamiltonian,
\begin{equation}
H=Q^\dagger Q+Q Q^\dagger=\hbar\omega\left[a^\dagger a+b^\dagger b\right]=
\hbar\omega\left[z\frac{\partial }{\partial z}
+\theta\frac{\partial}{\partial\theta}\right]\ .
\end{equation}

These operators thus act on wave functions $\psi(z,\theta)$.
Because of the Grassmann property $\theta^2=0$, a power series
expansion of such a function terminates at a finite order, in the
present case at first order since only one $\theta$ variable is involved,
\begin{equation}
\psi(z,\theta)=\psi_B(z)+\theta\psi_F(z)\ \ ,\ \ 
\psi_F(z)=\frac{\partial}{\partial\theta}\psi(z,\theta)\ ,
\end{equation}
where, assuming that $\psi(z,\theta)$ itself is Grassmann even,
the bosonic component $\psi_B(z)$ is Grassmann even while the
fermionic one $\psi_F(z)$ is Grassmann odd, as it should considering
the analogous structure of the space of quantum states.
In particular, the general wave function representing the
energy eigenstates $|n,0\rangle$ and $|n-1,1\rangle$
with value $E(n)=\hbar\omega n$ is given as
\begin{equation}
\psi_n(z,\theta)=B_n\frac{z^n}{\sqrt{n!}}\,+\,
F_n\theta\frac{z^{n-1}}{\sqrt{(n-1)!}}\ ,
\end{equation}
where $B_n$ and $F_n$ are arbitrary phase factors
associated to the bosonic and fermionic components of this
wave function.

The supersymmetry charges $Q$ and $Q^\dagger$ act on such general
wave functions as
\begin{equation}
Q\psi(z,\theta)=\sqrt{\hbar\omega}\ z\psi_F(z)\ \ ,\ \ 
Q^\dagger\psi(z,\theta)=\sqrt{\hbar\omega}\ \theta\partial_z\psi_B(z)\ .
\end{equation}
Thus introducing a complex valued Grassmann odd constant parameter
$\epsilon$ associated to the symmetries generated by the supercharges
$Q$ and $Q^\dagger$, one has for the general
self-adjoint combination of supercharges
\begin{equation}
Q_\epsilon=\epsilon Q+Q^\dagger\epsilon^\dagger=
\epsilon Q-\epsilon^\dagger Q^\dagger\ ,
\end{equation}
the action
\begin{equation}
Q_\epsilon\psi(z,\theta)=\sqrt{\hbar\omega}
\left[\left(z\epsilon\psi_F(z)\right)\ +\
\theta\left(\epsilon^\dagger\partial_z\psi_B(z)\right)\right]\ .
\end{equation}
Consequently, given the variations 
$\delta_\epsilon\psi(z,\theta)=iQ_\epsilon\psi(z,\theta)$, the bosonic
and fermionic components of such wave functions are transformed according
to the rules
\begin{equation}
\delta_\epsilon\psi_B(z)=i\sqrt{\hbar\omega}\,z\epsilon\psi_F(z)
\ \ ,\ \ 
\delta_\epsilon\psi_F(z)=i\sqrt{\hbar\omega}\,
\epsilon^\dagger\partial_z\psi_B(x)\ .
\label{eq:SUSYvariation1}
\end{equation}
These expressions thus provide the infinitesimal supersymmetry transformations
of the wave functions of the system. We shall come back to these
relations hereafter.

In order to identify which type of classical system corresponds
to the present situation, let us now introduce the configuration and
momentum space degrees of freedom through the usual relations,\cite{GovCOPRO2}
\begin{equation}
\begin{array}{r c l}
a=\sqrt{\frac{m\omega}{2\hbar}}\left[x+\frac{i}{m\omega}p\right]\ \ &,&\ \ 
a^\dagger=\sqrt{\frac{m\omega}{2\hbar}}\left[x-\frac{i}{m\omega}p\right]\ ,\\
 & & \\
b=\sqrt{\frac{m\omega}{2\hbar}}\left[\theta_1+i\theta_2\right]\ \ &,&\ \ 
b^\dagger=\sqrt{\frac{m\omega}{2\hbar}}\left[\theta_1-i\theta_2\right]\ .
\end{array}
\end{equation}
Note well that the variables $x$, $p$, $\theta_1$ and $\theta_2$, which
are assumed to be self-adjoint, $x^\dagger=x$, $p^\dagger=p$,
$\theta^\dagger_1=\theta_1$, $\theta^\dagger_2=\theta_2$, are still
operators at this stage. The decomposition of the fermionic operators
$b$ and $b^\dagger$ in these terms is of course to maintain as manifest
as possible the parallel between the bosonic and fermionic sectors of
the system, which are exchanged under supersymmetry transformations.
Given these operator redefinitions, it follows that the only nonvanishing
(anti)commutators are (note that the operators $\theta_1$ and $\theta_2$
thus anticommute with one another, $\left\{\theta_1,\theta_2\right\}=0$)
\begin{equation}
[x,p]=i\hbar\ \ ,\ \ 
\left\{\theta_1,\theta_1\right\}=\frac{\hbar}{m\omega}=
\left\{\theta_2,\theta_2\right\}\ .
\label{eq:quantumbrackets}
\end{equation}
Furthermore, the Hamiltonian operators is then expressed as
\begin{equation}
H=\frac{p^2}{2m}+\frac{1}{2}m\omega^2x^2+im\omega^2\theta_1\theta_2\ ,
\label{eq:Hamiltonian}
\end{equation}
leading to the operator equations of motion in the Heisenberg picture,
\begin{equation}
\begin{array}{l c l}
i\hbar\dot{x}=\left[x,H\right]=i\hbar\frac{p}{m}\ \ \ &,&\ \ \ 
i\hbar\dot{p}=\left[p,H\right]=-i\hbar m\omega^2x\ ,\\
 & & \\ 
i\hbar\dot{\theta}_1=\left[\theta_1,H\right]=i\hbar\omega\theta_2\ \ \ &,&\ \ \ 
i\hbar\dot{\theta}_2=\left[\theta_2,H\right]=-i\hbar\omega\theta_1\ .
\end{array}
\label{eq:quantumEM}
\end{equation}
It is also possible to determine how the supercharges $Q$ and $Q^\dagger$\
act on the operators $x$, $p$, $\theta_1$ and $\theta_2$, an exercise left
to the reader (of which the results are used hereafter).

Through the correspondence principle, the (anti)commutation relations
(\ref{eq:quantumbrackets}) are required to translate into the following 
classical Grassmann graded Poisson bra\-ckets for the associated degrees 
of freedom,
\begin{equation}
\left\{x,p\right\}=1\ \ ,\ \ 
\left\{\theta_1,\theta_1\right\}=-\frac{i}{m\omega}=
\left\{\theta_2,\theta_2\right\}\ ,
\end{equation}
with now all the variables $x$, $p$ , $\theta_1$ and $\theta_2$ 
real under complex conjugation, $x$ and $p$ being ordinary commuting
Grassmann even degrees of freedom, but $\theta_1$ and $\theta_2$ 
being anticommuting Grassmann odd degrees of freedom
associated to the fermionic sector of the system. At the classical level, 
the Hamiltonian is given by the same expression as in (\ref{eq:Hamiltonian}). 
In particular, using these Grassmann graded Poisson brackets, 
at the classical level the same Hamiltonian equations of motion 
are recovered as those in (\ref{eq:quantumEM}) for the quantum operators. 
These classical equations of motion follow through the variational 
principle from the first-order Hamiltonian action
\begin{equation}
S[x,p,\theta_1,\theta_2]=\int dt\left\{
\frac{1}{2}\left[\dot{x}p-\dot{p}x\right]
-\frac{1}{2}im\omega\left[\dot{\theta}_1\theta_1+\dot{\theta}_2\theta_2\right]
-H\right\}\ .
\end{equation}
 
Using the Hamiltonian equation of motion for $x$ in order to
reduce its conjugate momentum $p$, namely $p=m\dot{x}$, and also
introducing the complex valued Grassmann odd variable
$\theta=\theta_1+i\theta_2$, it then follows finally that the Lagrange function
of the system is given by,\footnote{Note that up to a total time
derivative term this function is indeed real under complex conjugation, 
because of the Grassmann odd character of the fermionic degree of freedom 
$\theta(t)$. Some total
derivative terms in time have been ignored to reach this expression,
and to bring it into such a form that no time derivatives of order
strictly larger than unity appear in the action.}
\begin{equation}
L=\frac{1}{2}m\dot{x}^2-\frac{1}{2}m\omega^2x^2
+\frac{1}{2}im\omega\theta^\dagger\dot{\theta}-
\frac{1}{2}m\omega^2\theta^\dagger\theta\ .
\label{eq:SUSYL}
\end{equation}
{}From the above considerations, it should then follow that the
transformations associated to the supercharges $Q$ and $Q^\dagger$
generate global symmetries of this action. These
transformations are given by\footnote{Compared to the previous
parametrisation, a factor $(-\sqrt{\hbar\omega}/2)$ has been absorbed
into the normalisation of the supersymmetry constant parameters
$\epsilon$ and $\epsilon^\dagger$ or supercharges $Q$ and $Q^\dagger$. 
Note also that these expressions
are consistent with the properties under complex conjugation of
the different degrees of freedom as well as their Grassmann parity.}
\begin{equation}
\delta_Q x=i\epsilon\theta+i\epsilon^\dagger\theta^\dagger\ \ ,\ \ 
\delta_Q\theta=2i\epsilon^\dagger\left(x+\frac{i}{\omega}\dot{x}\right)\ \ ,\ \ 
\delta_Q\theta^\dagger=-2i\epsilon
\left(x-\frac{i}{\omega}\dot{x}\right)\ .
\label{eq:SUSYvariation2}
\end{equation}
And indeed, it may readily be checked that the infinitesimal variation
of the Lagrange function (\ref{eq:SUSYL}) then reduces to
a simple total time derivative, thus establishing the supersymmetry
invariance of this system also at the classical level. Applying
Noether's general analysis to this new type of symmetry
for which the parameters are Grassmann odd quantities, leads back
to the conserved supercharges generating these 
transformations.\footnote{In such an analysis, one should beware
of the surface terms induced by the supersymmetry transformation applied
to the action, which also contribute to the definition of the
Noether charges.\cite{GovBook}}

Once again, a few general lessons may be drawn from the above
considerations, which remain valid in the case also of supersymmetric
field theories. The supercharges $Q$ and $Q^\dagger$ define transformations
between the states $|n,0\rangle$ and $|n-1,1\rangle$, except for the
vacuum state $|n=0,0\rangle$ which remains invariant under
supersymmetry. Hence, all these pairs of states for $n\ge 1$ define
2-dimensional supermultiplets, namely irreducible representations of
the supersymmetry algebra, combining a bosonic and a fermionic state
degenerate in energy. In the holomorphic wave function
representation, the bosonic component is given by 
$\psi(z,\theta)=\psi_B(z)=z^n/\sqrt{n!}$, and the fermionic one by 
$\psi(z,\theta)=\theta\psi_F(z)=\theta z^{n-1}/\sqrt{(n-1)!}$. 
{}From a certain point of view, the bosonic phase space $z$ of the system
has been extended into a super-phase space of degrees of freedom
$(z,\theta)$, on which supersymmetry transformations act through
$Q=\sqrt{\hbar\omega}z\partial_\theta$ and 
$Q^\dagger=\sqrt{\hbar\omega}\theta\partial_z$, thereby inducing
a map between the bosonic and fermionic components of a general
super-phase space wave function $\psi(z,\theta)$, according to the
rules in (\ref{eq:SUSYvariation1}). In particular, note that the
lowest component $\psi_B(z)$ of such a super-wave function is mapped
into its fermionic component, while its highest component $\psi_F(z)$
is mapped into the $z$-derivative of its lowest component.
If one recalls that through a process of second-quantisation, quantum
fields may be seen, in a certain sense,
to correspond to quantum mechanical wave functions
of 1-particle states, this remark suggests that by extension, supersymmetric
quantum field theories should be constructed in terms of superfields
depending not only on the usual spacetime coordinates $x^\mu$, but also
on some collection of Grassmann odd variables, in order to extend
usual Minkowski spacetime into some form of a superspace\cite{SS} Minkowski
spacetime, all in a manner consistent with the Poincar\'e covariance
properties required of such theories. Consequently, these Grassmann odd
variables must be chosen to determine specific spinor representations
of the Lorentz group, namely right- and left-handed Weyl spinors
$\theta_\alpha$ and $\overline{\theta}^{\dot{\alpha}}$.

Indeed, when developing this point of view, it appears that
such superfields decompose into specific bosonic and fermionic
components, defining a supermultiplet, with supersymmetry transformations
mapping these components into one another. In particular, the transformation
of the highest component always includes the spacetime derivative of some
of the lower components.

{}From the field theory point of view however, it would be more appropriate
to develop the same considerations rather in terms of the system degrees of
freedom $x(t)$ and $\theta(t)$, which are transformed into one another
as shown in (\ref{eq:SUSYvariation2}). Indeed, as argued in
Sec.~\ref{Sec2}, fields may be viewed as collections of oscillators
fixed at all points in space, namely $\phi(t,\vec{x}\,)=x_{\vec{x}}(t)$ and
$\psi(t,\vec{x}\,)=\theta_{\vec{x}}(t)$ for scalar and spinor fields,
respectively, and coupled to one another through
their spatial gradients in order to ensure spacetime Poincar\'e and
Lorentz invariance. In the case of the present simple supersymmetric mechanical
model, these bosonic and fermionic degrees of freedom $x(t)$ and $\theta(t)$
thus define a certain type of ``field'' supermultiplet (rather than a
supermultiplet of quantum states, as in the discussion above), of which
the variation of its highest component includes again the time derivative of 
some of its lower components. Upon quantisation, these transformation 
properties also apply to the quantum operators, and translate into 
the specific transformation rules for the quantum states described above. 
When extended to field theories, these features survive, this time 
in terms of bosonic and fermionic field degrees of freedom. Note 
that in (\ref{eq:SUSYL}), time derivatives of bosonic degrees of 
freedom contribute in quadratic form to the Lagrange function, whereas 
for fermionic ones they contribute to linear order. This fact, 
when extended to a relativistic framework, remains valid as well. 
The Klein--Gordon Lagrangian density is quadratic in spacetime gradients 
of the scalar field, but the Dirac Lagrangian density is linear in 
such derivatives of the Dirac spinor. For reasons explained previously, 
these features are necessary for a consistent dynamics of Grassmann even 
and odd, or integer and half-integer spin degrees of freedom.

For the sake of completing the discussion of the present simple supersymmetric
quantum mechanical model of harmonic oscillators, let us indeed show how
a superfield calculus may be developed already in this case. Again, 
the general lessons following from such an approach readily 
extend to the superfield constructions of supersymmetric field theories
in which the constraints of spacetime Poincar\'e covariance are then
also accounted for through the knowledge of the
representation theory of the Lorentz group.

The Hamiltonian of any given system is the generator of translations in time.
Given that the anticommutator of supercharges produces the Hamiltonian,
as shown for example in (\ref{eq:SUSYalgebra}), means that supersymmetry
transformations correspond to taking some sort of square-root of
translations in (space)time, in a new ``dimension'' of (space)time
which must be parametrised by a Grassmann odd coordinate this time,
since supercharges map bosonic and fermionic states into one another.
In addition to the bosonic time coordinate $t$, let us thus extend
time into a ``supertime'' by also introducing a complex valued Grassmann
odd coordinate, which shall be denoted $\eta$ (the customary
notation $\theta$ for Grassmann odd superspace coordinates being already
used for the fermionic degrees of freedom of the system, $\theta(t)$), 
and its complex conjugate $\eta^\dagger$. Thus we now have the superspace
(or rather for this mechanical model simply ``supertime'') spanned
by the coordinates $(t,\eta,\eta^\dagger)$. Supersymmetry transformations
generated by $Q$ and $Q^\dagger$ should then correspond to translations
in the Grassmann odd directions in superspace, in the same way that
transformations in time generated by the Hamiltonian correspond to
translations in the Grassmann even direction of superspace. By analogy
with the operator $i\partial_t$ generating the latter translations, and
representing the action of the Hamiltonian on degrees of freedom, the
naive choice for the supercharges would be $Q=-i\partial_\eta$ and
$Q^\dagger=i\partial_{\eta^\dagger}$. However, a quick check then finds
that all anticommutators $\left\{Q,Q,\right\}$, 
$\left\{Q^\dagger,Q^\dagger\right\}$ and $\left\{Q,Q^\dagger\right\}$
vanish, thus not reproducing the supersymmetry algebra 
in (\ref{eq:SUSYalgebra}). Hence, in order that the anticommutator 
of $Q$ and $Q^\dagger$ also reproduces the Hamiltonian, it is necessary 
that while a translation is performed in $\eta$ and $\eta^\dagger$, 
a translation in $t$ be also included in an amount proportional 
to the Grassmann odd coordinates in superspace. It turns out that 
an appropriate choice is given by\footnote{Some properties have to be 
met in the whole construction, such as preserving under supersymmetry
transformations the real character under complex conjugation of 
the superfield considered hereafter. This leaves open a series of
possible choices, essentially related to possible phase factors in the
combinations defining the superspace differential operators introduced
hereafter.}
\begin{equation}
Q=-i\partial_\eta+\frac{2}{\omega}\eta^\dagger\,\partial_t\ \ \ ,\ \ \ 
Q^\dagger=i\partial_{\eta^\dagger}-\frac{2}{\omega}\eta\,\partial_t\ .
\end{equation}
A direct calculation finds that these operators obey the supersymmetry
algebra
\begin{equation}
\left\{Q,Q\right\}=0=\left\{Q^\dagger,Q^\dagger\right\}\ \ ,\ \ 
\left\{Q,Q^\dagger\right\}=\left(-\frac{2}{\sqrt{\hbar\omega}}\right)^2\,
\left(i\hbar\partial_t\right)\ ,
\end{equation}
in perfect correspondence with the abstract algebra in (\ref{eq:SUSYalgebra})
(one should recall that a rescaling by a factor $(-\sqrt{\hbar\omega}/2)$
of the supersymmetry pa\-ra\-me\-ters $\epsilon$ and $\epsilon^\dagger$ 
or the supercharges $Q$ and $Q^\dagger$ has been applied in the 
intervening discussion).

In order to readily construct manifestly supersymmetric invariant Lagrange
functions, it proves necessary to also use another pair of superspace
differential operators, that anticommute with the supercharges, and
define so-called superspace covariant derivatives. These supercovariant
derivatives thus enable one to take derivatives of superfields in a manner
consistent with supersymmetry transformations. Again, a convenient
choice turns out to be
\begin{equation}
D=\partial_\eta-\frac{2i}{\omega}\eta^\dagger\,\partial_t\ \ \ ,\ \ \ 
D^\dagger=-\partial_{\eta^\dagger}+\frac{2i}{\omega}\eta\,\partial_t\ ,
\end{equation}
leading to the algebra
\begin{equation}
\left\{D,D\right\}=0=\left\{D^\dagger,D^\dagger\right\}\ \ ,\ \ 
\left\{D,D^\dagger\right\}=\left(-\frac{2}{\sqrt{\hbar\omega}}\right)^2\,
\left(i\hbar\partial_t\right)\ ,
\end{equation}
as well as the required properties
\begin{equation}
\left\{Q,D\right\}=0\ ,\
\left\{Q,D^\dagger\right\}=0\ ,\
\left\{Q^\dagger,D\right\}=0\ ,\
\left\{Q^\dagger,D^\dagger\right\}=0\ .
\end{equation}

Consider now an arbitrary Grassmann even superfield on superspace, 
namely a function $X(t,\eta,\eta^\dagger)$. Without loss of generality 
(by distinguishing its real and imaginary parts), it is always possible
to assume that such a superfield obeys a reality condition,
\begin{equation}
X^\dagger(t,\eta,\eta^\dagger)=X(t,\eta,\eta^\dagger)\ .
\end{equation}
On account of the Grassmann odd character of the coordinate $\eta$,
namely the fact that $\eta^2=0={\eta^\dagger}^2$, the general form
of such a real superfield is given by
\begin{equation}
X(t,\eta,\eta^\dagger)=x(t)+i\eta\theta(t)+i\eta^\dagger\theta^\dagger(t)+
\eta^\dagger\eta\,f(t)\ ,
\end{equation}
where $x(t)$ and $f(t)$ are real bosonic degrees of freedom,
whereas $\theta(t)$ and $\theta^\dagger(t)$ are complex valued fermionic
ones, complex conjugates of one another. Indeed, it will turn out that
$x(t)$ and $\theta(t)$ correspond to the degrees of freedom considered
above, while $f(t)$ will be seen to be simply an auxiliary degree of
freedom without dynamics, whose equation of motion is purely algebraic
and such that upon its reduction the system described in (\ref{eq:SUSYL})
is recovered. This is a generic feature of superfields in supersymmetric
field theories: they include auxiliary fields which are reduced through
their algebraic equations of motion. However, in the superspace formulation,
there are required for a supersymmetric covariant superspace calculus.

These choices having been specified, it is now straightforward to
es\-ta\-blish how the different components $(x,\theta,\theta^\dagger,f)$ 
(namely, the components of the terms in $1$, $i\eta$, $i\eta^\dagger$ and 
$\eta^\dagger\eta$ in the $\eta$-expansion of superfields) of
real superfields transform under supersymmetry transformations. By
considering the explicit evaluation of
\begin{equation}
\delta_Q\,X=i\left[\epsilon Q-\epsilon^\dagger Q^\dagger\right]\,X\ ,
\end{equation}
$\epsilon$ and $\epsilon^\dagger$ being the arbitrary complex valued
Grassmann odd constant supersymmetry parameters, complex conjugates of 
one another, it readily follows that the components vary according to
\begin{equation}
\begin{array}{r c l}
\delta_Q x=i\epsilon\theta+i\epsilon^\dagger\theta^\dagger\ \ &,&\ \ 
\delta_Q \theta=i\epsilon^\dagger\left[f+\frac{2i}{\omega}\dot{x}\right]\ ,\\
 & & \\
\delta_Q \theta^\dagger=-i\epsilon\left[f-\frac{2i}{\omega}\dot{x}\right]
\ \ &,&\ \
\delta_Q f=-\frac{2}{\omega}
\left[\epsilon\dot{\theta}-\epsilon^\dagger\dot{\theta}^\dagger\right]\ .
\end{array}
\label{eq:SUSYvariation3}
\end{equation}
It is of interest to compare these transformation rules to those
given in (\ref{eq:SUSYvariation2}).

Here appears yet another generic feature of the superfield technique.
One notices that the highest component $f(t)$ in $\eta^\dagger\eta$ of the 
superfield $X(t,\eta,\eta^\dagger)$ transforms under supersymmetry
as a total derivative in time. In the context of supersymmetric 
field theories, the highest component of superfields transforms as 
a total spacetime divergence. Thus, if one chooses for the Lagrange 
function or Lagrangian density the highest component of any relevant
superfield, under any supersymmetry transformation the action of the system 
is invariant up to a total derivative, thus indeed defining an invariance 
of its equations of motion. In superspace, supersymmetric invariant 
actions are given by the highest component of superfields, in our case
the $\eta^\dagger\eta$ component,
\begin{equation}
S[X]=\int dt\,d\eta\,d\eta^\dagger\,F(X)\ ,
\end{equation}
where $F(X)$ is any real valued superfield constructed out of the
basic superfield $X$ and its derivatives obtained through the action
of the supercovariant derivatives $D$ and $D^\dagger$. In this
expression, the definition of Grassmann integration is such that
\begin{equation}
\int\,d\eta\,d\eta^\dagger\,1=0\ \ ,\ \ 
\int\,d\eta\,d\eta^\dagger\,\eta=0\ \ ,\ \ 
\int\,d\eta\,d\eta^\dagger\,\eta^\dagger=0\ \ ,\ \ 
\int\,d\eta\,d\eta^\dagger\,\eta^\dagger\eta=1\ \ ,\ \ 
\end{equation}
while the result for any linear combination of these $\eta$-monomials
is given by the appropriate linear combination of the resulting 
integrations (the usual integral over Grassmann even variables being 
also linear for polynomials).

It turns out that the choice corresponding to the supersymmetric harmonic
oscillator in (\ref{eq:SUSYL}) is given by (one has, by construction
of the supercovariant derivatives, $(DX)^\dagger=D^\dagger X$ for the
real superfield $X$)
\begin{equation}
S[X]=\int dt\,d\eta\,d\eta^\dagger\,\left[
-\frac{1}{8}m\omega^2\left(D^\dagger X\right)\left(DX\right)\,-\,
\frac{1}{4}m\omega^2X^2\right]\ .
\end{equation}
Working out the superspace components of this expression, it reduces to
\begin{equation}
\begin{array}{r l}
S[x,\theta,\theta^\dagger,f]=\int dt\,\Big\{&
\frac{1}{8}m\omega^2\left[f^2+\frac{4}{\omega^2}\dot{x}^2+
\frac{2i}{\omega}\left(\theta^\dagger\dot{\theta}+
\theta\dot{\theta}^\dagger\right)\right]\,-\,\\
 & \\
&-\frac{1}{2}m\omega^2\left(fx+\theta^\dagger\theta\right)\Big\}\ .
\end{array}
\end{equation}
Since no time derivatives of the highest superfield component $f(t)$ 
contribute to this action, this degree of freedom is indeed auxiliary
with a purely algebraic equation of motion given by
\begin{equation}
f(t)=2x(t)\ .
\end{equation}
Upon reduction of this auxiliary degree of freedom, one recovers
precisely the Lagrange function in (\ref{eq:SUSYL}), up to a total
derivative in time $d/dt(-im\omega\theta^\dagger\theta/4)$, while for
the remaining dynamical degrees of freedom $x(t)$, $\theta(t)$ and
$\theta^\dagger(t)$, the supersymmetry transformations 
(\ref{eq:SUSYvariation3}) coincide then exactly with those in
(\ref{eq:SUSYvariation2}).

Having achieved the construction of the harmonic oscillator
with a single supersymmetry generator ${\mathcal N}=1$ from these different
but complementary points of view, one may
wonder whether generalisations to types of potentials other
than the quadratic one in $X^2$, to more general dynamics,
and for a larger number $\mathcal N$ of supersymmetries, are possible. 
The interested reader is invited to explore such issues further, which have been
addressed in the literature already to a certain extent.\cite{SUSYQM}

We have thus shown how a superspace extension of the time coordinate
into superspace coordinates $(t,\eta,\eta^\dagger)$ over which
a superspace calculus is defined for superfields, readily allows for
a systematic approach to the construction of supersymmetric quantum
mechanical models. This superspace calculus displays already the
features generic to the superspace techniques of superfields for the
construction of supersymmetric invariant field theories in
Minkowski spacetime. In the latter cas, it is spacetime itself which
is extended into a ``superspacetime'' 
$(x^\mu,\theta_\alpha,\overline{\theta}_{\dot{\alpha}})$ of
bosonic and Weyl spinor coordinates, the latter appearing with
multiplicities depending on the number $\mathcal N$ of supersymmetries
acting on the theory.\cite{Deren,SS}

\section{An Invitation to Superspace Exploration}
\label{Sec5}

As recalled in Sec.~\ref{Sec2}, by enforcing Poincar\'e invariance,
the ordinary bosonic harmonic oscillator extends naturally into
the quantum field theory of a scalar field describing relativistic
quantum point-particles of zero spin. One could attempt pursuing the
same road starting from the fermionic harmonic oscillator described
above and reach again the Dirac or the Majorana equation for spin 1/2 
charged or neutral particles, but the task would be quite much more 
involved, since the answer is known to require a 4-component
complex valued field, the parity invariant spinor representation of the
Lorentz algebra. Rather, it is by considering the detailed
representation theory of the Lorentz group that the correct
answer is readily identified in simple algebraic terms.

Likewise, one could attempt to extend the simple $\mathcal N=1$
supersymmetric harmonic oscillator model into a relativistic invariant
quantum field theory, which should thus include both a scalar field and
a Dirac--Majorana spinor. It is indeed possible to construct by hand such
a field theory, known as the Wess-Zumino model,\cite{WZ} the simplest example
of a $\mathcal N=1$ supersymmetric quantum field theory in
4-dimensional Minkowski spacetime. However, the approach is very
much streamlined by expressing everything in terms of superfields
defined over some superspace which extends Minkowski spacetime
by including some further Grassmann odd coordinates corresponding
to specific Weyl spinors. This is the superspace construction of
$\mathcal N=1$ supersymmetric quantum field theories.\cite{SS} Truly a
quantum geometer's approach to a possible quantum geometry
of spacetime.

However once again, in order to classify and identify the realm
of possible supersymmetric quantum field theories, for whatever number
$\mathcal N$ of supersymmetries, and whatever dimension of Minkowski
spacetime, and even whatever type of interactions consistent with the
requirements of perturbative renormalisability,
a discussion based on the possible algebraic structures
merging and intertwining together the Poincar\'e algebra with 
Grassmann odd generators mapping bosons to fermions and vice-versa,
is the most efficient approach. It is no small feat that such a
complete and finite dimensional classification has been achieved.\cite{Haag}
It is a nontrivial fact that such solutions exist, and also that
they are only a small finite number of possibilities consistent with the
rules of quantum field theory, in par\-ti\-cu\-lar unitarity and causality.
Clearly, such a situation gives credence to the suggestion that
such a combination of Poincar\'e covariance and supersymmetry invariance
brings us onto the right track towards the quest for a final unification.

The usefulness, relevance, and even meaning, of these different
remarks should find a nice simple illustration with the previous
quantum me\-cha\-ni\-cal model. These different avenues towards
the construction and classification of
supersymmetric quantum field theories have been developed and
discussed during the actual lectures delivered at the Workshop. However,
since this material is widely available in the literature,\cite{WB} and
in much detailed form, while the lectures themselves were to a large
extent based on those of Ref.~\refcite{Deren}, we shall stop
short here from pursuing any further the discussion of such
field theories and their particle content, except for just one last
remark.

The supersymmetric field theories simplest to construct in
4-dimensional Minkowski spacetime involve a single supersymmetry
generator, $\mathcal N=1$, represented by a right-handed Weyl
spinor supercharge $Q_\alpha$ and its complex conjugate left-handed
Weyl spinor supercharge $\overline{Q}^{\dot{\alpha}}$, using the
dotted and undotted index notation (thus the single supercharge
combines into a single Majorana spinor). In this case, the
supersymmetry algebra is defined by the anticommutation relations,
\begin{equation}
\left\{Q_\alpha,Q_\beta\right\}=0=
\left\{\overline{Q}_{\dot{\alpha}},\overline{Q}_{\dot{\beta}}\right\}
\ \ \ ,\ \ \ 
\left\{Q_\alpha,\overline{Q}_{\dot{\beta}}\right\}=
2P_\mu\left(\sigma^\mu\right)_{\alpha\dot{\beta}}\ .
\end{equation}
Clearly, these relations are the natural extension to a Poincar\'e
covariant setting of the supersymmetry algebra in (\ref{eq:SUSYalgebra})
relevant to the quantum mechanical oscillator model. Indeed, the Hamiltonian
is the time component of the energy-momentum 4-vector $P^\mu$,
while the components of the Weyl supercharges $Q$ and $\overline{Q}$
all square to a vanishing operator, implying again important
cohomology properties in supersymmetric quantum field theories.
The Noether charge $P^\mu$ being also the generator for spacetime
translations, implies that in a certain sense supersymmetry transformations
correspond to taking the square-root of spacetime translations, requiring
spinor degrees of freedom for consistency with Lorentz covariance.

As mentioned in Ref.~\refcite{GovCOPRO2}, often one prefers,
if only for aesthetical reasons having to do with spacetime locality
and causality, to have a local or gauged symmetry as compared to a 
global symmetry acting identically and instantaneously throughout 
all of spacetime. Supersymmetry transformations as described
in these notes correspond to global symmetries. Indeed, their 
infinitesimal action generated by the supercharges and mapping 
bosonic and fermionic degrees of freedom into one another, 
involves arbitrary Grassmann odd parameters
which are spacetime independent constants. This situation suggests
that one should gauge supersymmetry transformations, namely
consider the possibility of constructing quantum field theories
invariant under the same types of transformations between their
degrees of freedom for which though, this time, the parameters are
local functions of spacetime. Since the anticommutator of supercharges
induces spacetime translations, it is clear that by gauging supersymmetry
one has to introduce new field degrees of freedom\cite{GovCOPRO2} --- the 
associated gauge fields, possessing both bosonic and fermionic degrees 
of freedom --- which are in direct relation to spacetime reparametrisations 
and local Lorentz transformations, the latter being precisely the local gauge
symmetries of general relativity for instance. In other words, in exactly
the same way that gauged internal symmetries lead to Yang--Mills interactions,
gauged supersymmetry implies the gravitational interaction through 
a dynamical spacetime metric field of helicity $\pm 2$ and its 
supersymmetric partner field, in fact a Majorana Rarita--Schwinger 
of helicity $\pm 3/2$. For this reason, gauged supersymmetric 
field theories are known as supergravity theories.\cite{WB,SUGRA} 
Such theories exist for spacetime dimensions ranging from $D=2$ to $D=11$. 
Again, it is no accident that M-theory, the modern nonperturbative
extension of superstring theory and a possible candidate for a final
unification yet to be constructed, exists only in a spacetime
of dimension $D=11$.\cite{Strings}

It is hoped that through the above analysis of a simple
su\-per\-sym\-me\-tric quantum mechanical model, the reader will
have understood enough of the general concepts and generic features
entering the formulation and the construction
of supersymmetric field theories, as well as of their
potential to address from novel and powerful points of view large fields
of pure mathematics itself, that he/she may feel sufficiently
secure in following the lead of such little white and precious
pebbles along the path, to embark on a journey of one's own 
onto the roads of the quantum geometer's superspaces, deep into 
the unchartered territories of supersymmetries and their yet to be 
discovered treasure troves in the eternally fascinating worlds of 
physics and mathematics, thereby fulfilling ever a little more,
this time with a definite African beat in the symphony,
humanity's unswaying and yet never ending quest for a complete
understanding of our destiny in the physical Universe, the eternal
yearning of man's soul.\cite{GovCOPRO2}

\section*{Acknowledgements}

The author acknowledges the Workshop participants for their many
constructive discussions and contributions on the matter of
these lectures. This work is partially supported by the 
Federal Office for Scientific, Technical and Cultural Affairs (Belgium) 
through the Interuniversity Attraction Pole P5/27.


\begin{thebibliography}{99}
\bibitem{Lambda}
For references on the cosmological constant problem, see,\\
J. Govaerts, {\sl The cosmological constant of one-dimensional matter
coupled quantum gravity is quantised\/}, contribution to this volume.
\end{thebibliography}
\end{document}